%% file: TOSEM.tex
\documentclass[acmsmall]{acmart}

\usepackage{amsmath,amssymb,amsfonts}
\usepackage{graphicx}
\usepackage{textcomp}
\usepackage{xcolor}
\usepackage{threeparttable}
\usepackage{graphics}%
\usepackage{multirow}%
\usepackage{amsthm}%
\usepackage{mathrsfs}%
\usepackage[title]{appendix}%
\usepackage{manyfoot}%
\usepackage{booktabs}%
\usepackage{algorithm}%
\usepackage{algorithmicx}%
\usepackage{algpseudocode}%
\usepackage{listings}%
\usepackage{url}
\usepackage[shortlabels]{enumitem}
\usepackage{float}
\usepackage{placeins}
\usepackage{booktabs} 
\usepackage{balance}
\usepackage{lipsum}

\usepackage{dsfont}
\usepackage{titlecaps}
\usepackage{array}
\usepackage{tabularx}
\usepackage{fancyvrb}

\usepackage{stmaryrd}
\usepackage{tikz, tcolorbox}
\usepackage{pgfplots}
\pgfplotsset{compat=1.18}
\usepackage{pgfplotstable}
\usepackage{makecell}
\usepackage{caption}
\usepackage{subcaption}
\usepackage{hyperref}
\usepgfplotslibrary{patchplots}
\usepgfplotslibrary{colorbrewer}
\usetikzlibrary{shadows}
\usetikzlibrary{arrows}
\usetikzlibrary{arrows.meta}
\usetikzlibrary{shadows.blur}

\tikzset{%
   tipSq/.tip={Square[open,fill=black]}
}

\pgfplotsset{cycle list/BrBG}

\newtheorem{definition}{Definition}

\DeclareMathOperator*{\argmax}{arg\,max}

\definecolor{prismgreen}{rgb}{0, 0.6, 0}
\definecolor{haiqprocblue}{rgb}{0, 0.0, 0.6}

\lstdefinelanguage{haiq}{
  keywords={%
      assert, pred, all, no, lone, one, some, check, run, but, let, implies, not, iff, in, and, or, set, sig, Int, int, if, then, else, exactly, disj, fact, fun, module, abstract, extends, open, none, univ, iden, seq, init, var, enum, formula, false, true, bool,
      },
        basicstyle=\color{black}\scriptsize\sffamily, 
        keywordstyle={\bfseries\color{black}},
        numberstyle=\tiny\color{black},
        belowcaptionskip=\baselineskip,
        comment=[l] {//}, 
        commentstyle= \color{prismgreen}, 
        tabsize=4, 
        captionpos=b, 
        escapechar=@, 
        belowcaptionskip=0em,
        belowskip=0em,
  }

\AtBeginDocument{%
  }

\setcopyright{acmlicensed}
\copyrightyear{2026}
\acmYear{2026}
\acmDOI{XXXXXXX.XXXXXXX}

\acmJournal{JACM}
\acmVolume{37}
\acmNumber{4}
\acmArticle{111}
\acmMonth{8}

\begin{document}

\title[Architecting Hybrid Quantum-Classical Software Systems]{Architecting Hybrid Quantum-Classical Software Systems: Exploration of the Design Trade-off Space with Quantitative Guarantees} 

\author{\'Alvaro M. Aparicio-Morales}
\email{amapamor@unex.es}
\orcid{0009-0009-5161-5498}

\author{Jose Garcia-Alonso}
\email{jgaralo@unex.es}
\orcid{0000-0002-6819-0299}

\author{Juan M. Murillo}
\email{juanmamu@unex.es}
\orcid{0000-0003-4961-4030}

\affiliation{%
  \institution{Universidad de Extremadura}
  \city{Cáceres}
  \state{Extremadura}
  \country{Spain}
}

\author{Javier C\'amara}
\email{jcamara@uma.es}
\orcid{0000-0001-6717-4775}

\affiliation{%
  \institution{ITIS Software}
  \city{Málaga}
  \state{Andalucía}
  \country{Spain}
}

\renewcommand{\shortauthors}{\'A.M. Aparicio-Morales et al.}

\begin{abstract}
     \textbf{Abstract.} Addressing problems beyond classical computing limits is sparking an increasing interest in Quantum Computing. However, despite their adequacy to address specific problems, quantum algorithms cover a limited subset of the functionality required in real-world computing systems. Additionally, they require expensive specialized hardware. To overcome this issue, hybrid (quantum-classical) software systems are emerging as a promising way to integrate both computing paradigms by applying the principles of Service-Oriented Architectures (SOA). Still, the design and deployment of hybrid service-based systems faces unique challenges like the idiosyncrasies and constraints of NISQ computers (e.g., algorithms that can only run in specific machines, disparate quality attribute metrics), and the management of structural and behavioural properties of service-based applications. From the SOA perspective, architectural decisions need to be made by performing a trade-off analysis and providing quantitative guarantees of system configurations under prescribed levels of uncertainty. In this paper, a method to explore the design space of quantum-classical applications is provided by a formalization of an architectural style of hybrid applications. The obtained results demonstrate that the proposed method successfully identifies decision boundaries. It enables the dynamic selection of the most suitable hybrid or classical configuration based on the user's QoS criteria.
\end{abstract}



\received{27 March 2026}
\received[revised]{DD Month YYYY}
\received[accepted]{DD Month YYYY}

\maketitle

\section{Introduction}
\label{sec:introduction}
\input{1_Introduction/introduction}

\section{Motivating Scenario}
\label{sec:motivating}
\input{2_MotivatingScenario/motivatingsecenario}

\section{Background}
\label{sec:background}
\input{3_Background/background}

\section{Design Trade-off Space Framework for Hybrid (Quantum-Classical) Software Systems}
\label{sec:approach}
\input{4_QCFramework/qcframework}

\section{Evaluation}
\label{sec:evaluation}
\input{5_Evaluation/evaluation}

\section{Threats to validity}
\label{sec:threats}
\input{6_ThreatsToValidity/threatstovalidity}

\section{Related Works}
\label{sec:related}
\input{7_RelatedWork/relatedwork}

\section{Conclusion and Future Work}
\label{sec:conclusion}
\input{8_ConclusionAndFutureWork/conclusionandfuturework}



\bibliographystyle{ACM-Reference-Format}
\bibliography{references}


\end{document}

%% file: 1_Introduction/introduction.tex
Quantum Computing (QC) brings the promise of a revolution in computing by providing capabilities to tackle problems that are either non-tractable or computationally prohibitive by today's ``classical'' computing standards. Such capabilities include e.g., prime number factorization \cite{shor1999polynomial} and simulation of quantum phenomena in nature, such as solving Navier-Stokes equations for weather forecasting~\cite{Gaitan2020}. Although QC is in the Noisy Intermediate-Scale Quantum (NISQ) era~\cite{DBLP:books/daglib/0046438}, novel solutions based on this technology are beginning to emerge~\cite{DBLP:journals/tits/StollenwerkOVMR20, DBLP:journals/sncs/DingCLSS21}. Therefore, it is expected that in the medium to long term, more quantum software will arise to tackle complex challenges~\cite{Murillo2024}.

However, despite its adequacy to address such specific classes of problem, QC covers only a narrow band of the spectrum of functionality needed to build real-world systems, which often require incorporating components to provide support for aspects such as logging, storage, and user interaction, to name a few. Furthermore, there are cases in which alternative quantum and classical implementations that cover the same functionality exist, and there are tradeoffs that make the choice of one alternative over the other more desirable, depending on the specific situation. Hence, the replacement of current classical components by their quantum counterparts results in systems with a hybrid (quantum-classical) workflow. This new type of hybrid software system must also operate under certain levels of trustworthiness and quality.

Beyond their narrow scope of functionality, quantum algorithms require specialized hardware that is costly to buy and maintain. Indeed, in-house quantum hardware ownership and management is justified only in a very limited number of cases~\cite{DBLP:journals/corr/abs-2404-11420}, and commoditization of quantum computing resources via platforms such as Amazon Braket (\url{https://aws.amazon.com/braket/}) or Azure Quantum (\url{https://quantum.microsoft.com/}) are progressively making quantum computing more affordable by providing it through models such as {\em Quantum Computing as a Service} ({\em QCaaS})~\cite{DBLP:journals/internet/Garcia-AlonsoRV22}.

As a result, service-based hybrid {\em quantum-classical} applications are naturally emerging as a model that enables integrating classical and quantum computing by applying the principles of Service-Oriented Architectures~\cite{Murillo2024}. 
However, to design quantum-classical applications that achieve a good balance among multiple non-functional properties, architects have to explore design spaces that are often poorly understood. 
Part of the intricacy in the design space stems from design decisions that often involve the selection and composition of loosely coupled, pre-existing components or services with different levels of quality (e.g., reliability, performance, cost) that may be offered by independent providers~\cite{mahdavi2013variability}. 
Moreover, beyond the constraints and the uncertainty that affects the behavior of the constituent components of classical service-based systems (e.g., faults, network delays, lack of control over third-party services residing in the cloud), quantum-classical service-based systems are subject to additional constraints that are derived from the incorporation of quantum services that can be deployed only on specific hardware~\cite{DBLP:conf/summersoc/SalmBBLWW20}, have disparate attributes and pricing models~\cite{10313627}, and are subject to additional sources of uncertainty (e.g., the output of a quantum circuit is generally not deterministic).

For this reason, architects need tools and techniques that can help them explore these complex design spaces and guide them to good designs. Providing such tool support requires investigating the following questions:

\begin{enumerate}[leftmargin=1.2cm,start=1,label={(\bfseries RQ\arabic*):}]
    \item How can we automatically identify the set of configurations that satisfy the architectural constraints characteristic of quantum-classical service-based systems? 
    \item How can we analyze the trade-offs among relevant non-functional system properties (e.g., cost, performance, reliability) in quantum-classical service-based systems across the architectural design space and under prescribed levels of uncertainty?
\end{enumerate}
Although there are existing approaches that enable the exploration of architectural design spaces under prescribed levels of uncertainty (e.g., C\'amara et al.~\cite{DBLP:journals/jss/CamaraGS19}), such proposals do not take into consideration some of the aspects that are unique to hybrid quantum-classical service-based systems. For this type of hybrid system, it is necessary to explicitly consider the characteristics of the hardware components, as well as the behaviour of the software, in order to identify the set of machines on which it can be executed. Likewise, it is necessary to establish a set of constraints for the service-machine relationship to ensure that the different deployment configurations are structurally correct. This combination of possible service-machine configurations implies that the service quality attributes are dependent on the machine with which they are related. In this paper, this gap have been addressed by contributing: (i)~a formalization of an architectural style for service-based quantum-classical applications, (ii)~a method that exploits this formalization to automatically generate system configurations and deployments with quantitative guarantees about their quality levels under prescribed levels of uncertainty, and (iii)~two case studies to evaluate our method that investigates our research questions. 

Our results demonstrate that the proposed method enables architectural decision-making for hybrid quantum-classical systems through the automated evaluation of the configuration space. Concretely, our approach allows for the identification of trade-offs among solutions, providing support for the selection of configurations that are appropriate to specific contexts of system use.

The rest of the paper is organized as follows: Section~\ref{sec:motivating} presents a scenario that motivates our proposal. Section~\ref{sec:background} presents some background on quantum computing and the formal analysis technique that supports our approach. Section~\ref{sec:approach} describes our formalization and method for analyzing service-based quantum-classical system architectures. Sections~\ref{sec:evaluation} and ~\ref{sec:threats} discuss evaluation and threats to validity, respectively. Section~\ref{sec:related} discusses related work. 
Section~\ref{sec:conclusion} presents some conclusions and lines for future work.

%% file: 2_MotivatingScenario/motivatingsecenario.tex
\begin{figure}[H]
    \centering
    \includegraphics[width=0.7\textwidth]{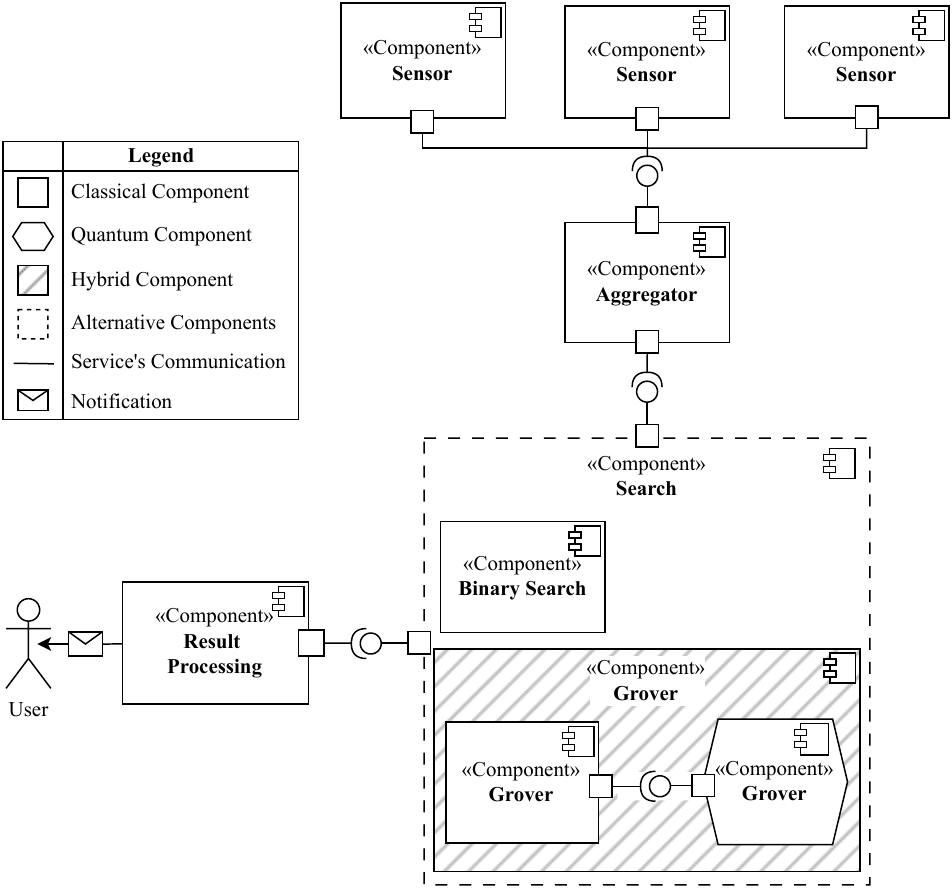}
    \Description{Architecture of Hybrid Search Application, contains three sensor services, connected to the aggregator service. The aggregator is also connected to the search service, either classical or hybrid. The search service is connected to the result processing service, which notifies the user.}
    \caption{Architecture of the HSA System}
    \label{fig:MotivatingScenario}
    
\end{figure}

We illustrate our approach on a Hybrid Search Application (HSA), a service-based application that aims at detecting anomalous measurements in a sensor network (Figure~\ref{fig:MotivatingScenario}). 
Sensors are exposed through services that send data upon request from an aggregator service, which merges the various sensor data streams and sends the aggregated data to a search service.
The search service can be implemented either as a classical service that runs, e.g. a binary search algorithm, or as a hybrid (quantum-classical) service in which search is performed using quantum computing through Grover's algorithm~\cite{grover1996fast}.
Finally, the result processing service is responsible for analysing the output of the search component and notifying the user of any possible anomaly.

Given the architecture of this application, alternative configurations present disparate levels of execution time, cost, reliability and quantum error in results that depend, among other factors, on the machines (either quantum or classical), where services are deployed (i.e., the same service implementation may experience quality variations across different machines). Likewise, in this motivating scenario, the uncertainties emerge from three sources. First, errors can be produced in the service execution at run time, emerging from unhandled exceptions, malicious attacks, etc. Second, cloud provider infrastructure can experience problems that can affect the availability of their machines where the application is deployed~\cite{7435330,6754595}. Finally, a third source of uncertainty is associated with errors in computational results that are produced by quantum computer noise~\cite {DBLP:books/daglib/0046438} and affect the accuracy of results.

In this context, finding an adequate architectural system design entails understanding the tradeoff space, identifying the set of configurations that satisfy: (i)~a set of structural constraints (e.g., sensor services must not directly communicate with the search service), (ii)~a set of behavioral properties (e.g., the system must eventually return a result), and (iii)~a set of quality requirements that can be expressed, e.g., as a combination of quantitative constraints and optimization objectives, such as the ones illustrated in Table~\ref{tab:qos}. Due to space reasons, the set of structural constraints and behavioural properties of the motivating scenario is collected in the Zenodo repository~\footnote{\href{https://zenodo.org/records/18482185?preview=1&token=eyJhbGciOiJIUzUxMiJ9.eyJpZCI6ImMwYmFlZjIzLTg2MDgtNGE3ZC05ZDkwLWNkZDAyZmU0YzA0ZCIsImRhdGEiOnt9LCJyYW5kb20iOiIwMDk3ZDYxNDcxOWRiOTQ0ZTIyZmM3YjkwZDhmMTk5MyJ9.ju-Ilwtn9bGmt7nyh_wnHCNCFSpZoUYRBpEB4zAx2Ln-qMv9_dJPjAE-KoS9sLCJ2QH7Hn2neohYHNdJdVQSXg}{Motivating Scenario Zenodo Link}}.

Generalizing this scenario, the problem we tackle is, given a set of constraints imposed by the architectural style of an application, as well as, a set of elements (service components, connectors, computational resources such as quantum and classical processing units where services can be deployed), and a set of requirements (quality, behavioral), identifying the set of system configurations that satisfy structural constraints, along with behavioral and quality requirements.

\begin{table}[H]
    \centering
    \caption{HSA - Quality of Service Requirements}
    \label{tab:qos}
    \begin{tabular}{c|l}
    \hline
        \textbf{Name} & \textbf{Description} \\
        \hline
        R1 & The average execution time must not exceed $t$ s/request. 
        \\
        \hline
        R2 & Execution cost must not exceed $e$ usd/request.
        \\
        \hline
        R3 & Failure rate must not exceed $r\%$.
        \\
        \hline
        R4 & Quantum Error must not exceed $q$ number of wrong executions/request.
        \\
        \hline
    \end{tabular}
\end{table}

%% file: 3_Background/background.tex
In this section, we first describe some context that concerns Quantum Computing, following with an introduction to the basics of the technique based on model finding and quantitative verification that we employ in our approach.

\subsection{Quantum Computing}
Quantum Computing is a paradigm based on leveraging quantum mechanics phenomena (e.g., superposition, entanglement) to perform computations using qubits, the fundamental units of quantum information. Superposition allows qubits to exist in a state where values are undetermined until the system collapses. Entanglement occurs when qubits become strongly correlated, such that the state of one cannot be described independently of the other~\cite{DBLP:books/daglib/0046438}.

The Quantum Processing Unit (QPU) is the core component where computation takes place. It consists of a set of qubits distributed on a quantum chip, where operations are typically performed between pairs of qubits. Furthermore, current hardware technology makes qubits highly sensitive to environmental noise. Consequently, QPUs often yield error-prone results. Given these limitations, and the fact that current qubit counts are insufficient for addressing large-scale problems, the field is currently in the Noisy NISQ era~\cite{DBLP:books/daglib/0046438}. Simultaneously, the lack of standardization in quantum hardware, combined with distinct problem-solving approaches, has led to the development of diverse QC technologies~\cite{DBLP:conf/summersoc/SalmBBLWW20}. The most widespread is the \textit{gate-based quantum computers}~\cite{10313627}; consequently, this type was selected for this study. It is considered the universal quantum computer~\cite{DBLP:books/sp/Hidary21}. Its operational model is analogous to classical CPUs, a set of logic gates (i.e., quantum gates) is applied to the bits (i.e., qubits) to perform computation. The arrangement of these gates on a set of qubits constitutes a quantum circuit. Key properties of a quantum circuit include its \textbf{\textit{width}}, defined as the number of qubits involved in the computation, and its \textbf{\textit{depth}}, defined as the longest path of sequential gates applied to a single qubit. Additionally, the number of \textbf{\textit{shots}} represents the number of times the circuit is executed to obtain a result. These three metrics are critical factors in determining whether a specific quantum circuit can be successfully mapped onto a given target QPU.

Moreover, current quantum hardware specification metrics are not the same for each quantum computer, and they depend on its provider. For instance, IBM offers a pricing model based on compute time per minute, whereas AWS Braket charges users based on the number of shots. Additionally, other metrics to evaluate performance on a quantum machine are appearing, such as Quantum Volume or CLOPS (Circuit Layer Operations per Second), making performance estimation different across platforms. In this study, the selected metric to evaluate the performance of a quantum machine is CLOPS because it was defined by IBM as a metric correlated with how fast a quantum processor can execute circuits.

The downside of these NISQ devices is that they require special conditions to perform computations optimally. As a result, they can only be housed in large infrastructures, such as data centres or research facilities owned by governments and major multinational corporations~\cite{DBLP:conf/wcre/AparicioMoralesHMGBSMM24}. Due to the stringent environmental and operational requirements for hosting quantum computers, most of them are accessible via cloud computing providers through their Software Development Kits (SDKs). This shift has enabled a new paradigm where these SDKs are used to develop classical services that act as entry points, by encapsulating the implementation of quantum algorithms in the cloud. This approach was proposed in~\cite{DBLP:journals/internet/Garcia-AlonsoRV22} and was referred to as a hybrid (quantum-classical) service. By leveraging these hybrid services, it becomes feasible to integrate them into existing software systems, providing an alternative to traditional computational services that perform in the same way. 
However, bringing this promise to fruition requires making practical considerations related to the way in which hybrid (quantum-classical) services can be integrated into real-world information systems and harmoniously co-exist with classical computing forming  hybrid quantum-classical software systems. In this context, ensuring trustworthiness in hybrid (quantum-classical) software is crucial to guarantee the reliability, security, and robustness of such integrated systems.
\subsection{Model-Finding with Quantitative Verification}
\label{sec:haiq}
In order to guarantee the structural validity of the generated design space while simultaneously assessing its quantitative characteristics, a framework that integrates structural constraints with stochastic behavioral properties is required. For this purpose, our proposal is underpinned by HaiQ~\cite{DBLP:conf/icse/Camara20}. This tool-supported approach synergizes probabilistic model checking~\cite{DBLP:conf/sfm/KwiatkowskaNP07} mechanisms with model-finding capabilities in the style of Alloy~\cite{DBLP:journals/cacm/Jackson19}, enabling the combined analysis of structural and behavioral properties in a quantitative fashion.
As input, HaiQ receives two specifications: (i)~a system model specified in a high level language (HaiQ) capable of capturing both structural and behavioral descriptions of system components, including probabilistic and other quantitative attributes (e.g., time, cost) and (ii)~a set of formalized requirements specified in a temporal logic language ({\em manifold probabilistic computation tree logic} or M-PCTL) that extends the PCTL logic commonly used in probabilistic model checkers to express (mostly quantitative) properties about collections of systems variants that can be generated from a HaiQ specification.
A {\sf HaiQ} model includes both a set of {\em signatures} and a set of variables. 
A {\em signature} is a type definition from which multiple instances or objects of that type can be generated to build a model of a system variant. 
The specification of each signature contains two parts: structural and behavioral. 
The structural part captures relations between the instances defined by the signatures, while the behavioral part expresses the (possibly stochastic) behavior of the instances. 
A HaiQ specification is formally characterized as a tuple $M = \langle B, \Sigma, \mathcal{C} \rangle$, where:
\begin{itemize}
    \item $B$ is a set of variables that range over finite integers or booleans. Variables can belong to a signature or be global.
    \item $\Sigma$ is a set of signatures that defines a hierarchy of types (including their behavior), where each signature $\sigma \in \Sigma$ is defined as a tuple $\langle \sigma^{\uparrow}, m, A, R \rangle$:
        \begin{itemize}
             \item $\sigma^{\uparrow}$ is a (possibly empty) parent signature from which $\sigma$ can inherit, meaning that it shares the same actions and variables as $\sigma^{\uparrow}$.
             \item $m$ is the multiplicity of the signature that prescribes how many instances of $\sigma$ can be generated. Allowed values are {\sf \small \textcolor{blue}{abstract}}, {\sf \small \textcolor{blue}{lone}}, {\sf \small \textcolor{blue}{one}}, {\sf \small \textcolor{blue}{some}}, {\sf \small \textcolor{blue}{set}}, which correspond to zero, at most one, exactly one, at least one, and any number of instances, respectively. 
            \item $A$ is a (possibly empty) set of actions $a = \langle g, \mathcal{D}_U \rangle$, where the action's guard $g$ is a boolean expression over the set of variables $B$, and $\mathcal{D}_U$ is a discrete probability distribution over a set of updates $U$. An update $u \in U$ assigns new values to one or more variables in $B$.
            \item $R$ is a set of relations with other signatures. A relation $r = \langle m_r, \sigma_r \rangle$ defines a relationship multiplicity  $m_r$ that takes values in the same range as $m$, and a signature $\sigma_r$ with which the relation is established.        
        \end{itemize}
    \item $\mathcal{C}$ is a set of structural constraints over the relations defined by signatures in $\Sigma$, expressed in first-order predicate logic. 
\end{itemize}

Probabilistic Computation Tree Logic (PCTL)~\cite{Hansson1994} is used to quantify properties related to probabilities and rewards in {\it single system specifications described as a probabilistic state machine} (e.g., discrete-time Markov chain -DTMC-, Markov decision process -MDP-, probabilistic timed automata or PTA). 
In contrast, M-PCTL targets {\it quantification across collections of design alternatives} that correspond, in this case, to the state machines generated from the set of architectural configurations that satisfy the constraints of a {\sf HaiQ} specification.
M-PCTL extends PCTL with checking of probability and reward-based properties to collections of models. Hence, quantification occurs over a pair $(\mathcal{M}, \rho)$, where $\mathcal{M}$ is a set of models, and $\rho$ is a set of reward functions.
M-PCTL includes three types of formula. Similarly to PCTL, it includes path ($\phi$) formulas (which are the same as in PCTL) and state ($\Phi$) formulas, but also an additional type of set formula ($\Psi$) that returns the collection of models that satisfy a particular quantitative constraint.
The syntax of M-PCTL is:
\begin{center}
$\Phi ::= {\tt true} | \; a \;  | \; \neg \Phi \; | \; \Phi \wedge \Phi \; |$ \\ 

${\sf someP}_{\sim pb} [\phi] \; | \; {\sf allP}_{\sim pb} [\phi] \; | \; {\sf maxP}_{\sim pb} [\phi] \; | \; {\sf minP}_{\sim pb} [\phi] \; |$ 

${\sf someR}^r_{\sim rb} [\phi] \; | \; {\sf allR}^r_{\sim rb} [\phi] \; | \; {\sf maxR}^r_{\sim rb} [\phi] \; | \; {\sf minR}^r_{\sim pb} [\phi]$

$\Psi ::= \; U \;  | \; \Psi \bigcup \Psi \; | \; \Psi \bigcap \Psi \; | \; \Psi^C \; |$

${\sf SsomeP}_{\sim pb} [\phi] \; | \; {\sf SallP}_{\sim pb} [\phi] \; | \; {\sf SmaxP} [\phi] \; | \; {\sf SminP} [\phi] \; |$ 

${\sf SsomeR}^r_{\sim rb} [\phi] \; | \; {\sf SallR}^r_{\sim rb} [\phi] \; | \; {\sf SmaxR}^r [\phi] \; | \; {\sf SminR}^r [\phi]$ 

\end{center}
 Concerning state formula quantifiers, ${\sf allP}$ and ${\sf someP}$ determine if the evaluation of $Pr(\phi)$ on all or some model in $\mathcal{M}$ satisfies $\sim pb$, whereas ${\sf maxP}$ determines if the maximum probability evaluated across elements of $\mathcal{M}$ satisfies $\sim pb$. 
We define their semantics as:
\begin{center}
$ \llbracket {\sf someP}_{\sim pb} [\phi] \rrbracket \equiv  \exists M \in \mathcal{M} : Pr_M(\phi) \sim pb $

$ \llbracket {\sf allP}_{\sim pb} [\phi] \rrbracket \equiv  \forall M \in \mathcal{M} : Pr_{M}(\phi) \sim pb $

$ \llbracket {\sf maxP}_{\sim pb} [\phi] \rrbracket \equiv  {\displaystyle \max_{M \in \mathcal{M}} Pr_{M}(\phi)} \sim pb$,
\end{center}
\noindent where  $Pr_{M}(\phi)$ denotes the evaluation of the probability $Pr(\phi)$ on model $M$. 
The analogous reward-based quantifiers ${\sf someR}^r$, ${\sf allR}^r$, ${\sf maxR}^r$, and ${\sf minR}^r$, are defined over the expected reward measure of PCTL, instead of the probabilistic one $Pr$ (c.f.~\cite{DBLP:conf/sfm/KwiatkowskaNP07}). 
The use of {\sf maxP/minP} and {\sf maxR/minR} quantifiers without a bound implies the quantification of the actual maximum/minimum probability or reward for the path formula $\phi$, e.g.:
$ \llbracket {\sf maxP} [\phi] \rrbracket \equiv {\displaystyle \max_{M \in \mathcal{M}} Pr_{M}(\phi)}$.
In set formulas, $U$ denotes the universe of models in $\mathcal M$ and $\Psi^C$ is the standard complement operator of set algebra. The semantics of the main quantifiers in set formulas is:
\begin{center}
$ \llbracket {\sf SallP}_{\sim pb} [\phi] \rrbracket \equiv  \{ M : \mathcal{M} \; | \; Pr_{M}(\phi) \sim pb \}$

$ \llbracket {\sf SmaxP} [\phi] \rrbracket \equiv  {\displaystyle \argmax_{M \in \mathcal{M}} Pr_{M}(\phi)}$
\end{center}
Quantifier ${\sf SsomeP}$ returns a singleton with an element drawn nondeterministically from ${\sf SallP}_{\sim pb} [\phi]$ if the set is not empty, and $\emptyset$ otherwise.

The previously defined formal model is concretely captured within the HaiQ specification language. As illustrated in \textit{Fig.}~\ref{haiq:signatureexample}, which depicts a portion of the motivating scenario, each formal component has a direct syntactic counterpart. This snippet allows for a detailed examination of how the theoretical definitions are instantiated. First, the bound $B$ is represented by the declared constant \textit{MAX\_TIMEOUT}. The set of signatures $\Sigma$ is instantiated by the \textit{PU} and \textit{QPU} signatures. 
Specifically, the inheritance $\sigma^{\uparrow}$ corresponds to the \textit{extends PU} of the \textit{QPU} signature. The multiplicity $m$ in the formal model is denoted by the keyword {\sf \small \textcolor{blue}{abstract}}. Furthermore, the set of actions $A$ can be observed in the line \textit{[services:qpair] true $\rightarrow$ (finished'=true);}. Regarding the relationships $R$, these are represented within the \textit{PU} signature, which contains the set of related services. 
Constraints $C$ are showcased in Listing~\ref{haiq:constraintexample}, while the M-PCTL properties are defined in Listing~\ref{haiq:propertiesexample}.

\begin{figure}[ht]
\caption{HaiQ Signature Examples}
\Description{HaiQ Signature examples}
\label{haiq:signatureexample}
\begin{minipage}{0.46\textwidth}
\begin{lstlisting}[language=haiq, breaklines=true, basicstyle=\scriptsize\ttfamily]
const MAX_TIMEOUTS;
abstract sig PU {services : some Service}
</
    formula cpulogicalperformancefactor;
    formula cpuram;
    formula cpubandwidth;
    formula cpucostfactor;
    formula qpuprize;
    formula clops;
    formula readoutminerror;
    formula readoutmaxerror;
    formula cpuerrorrate;
/>
\end{lstlisting}
\end{minipage}
\hfill
\begin{minipage}{0.5\textwidth}
\begin{lstlisting}[language=haiq, breaklines=true, basicstyle=\scriptsize\ttfamily]
abstract sig QPU extends PU {}
</
    formula cpulogicalperformancefactor;
    formula cpuram;
    formula cpubandwidth;
    formula cpucostfactor;
    formula qpuprize;
    formula clops;
    formula readoutminerror;
    formula readoutmaxerror;
    formula cpuerrorrate;
    var finished : bool init false;
    [services:qpair] true -> (finished'=true);
/>
\end{lstlisting}
\end{minipage}
\end{figure}

\begin{figure}[ht]
\Description{Representation in HaiQ syntax of a constraint and declaration of properties to study}
\begin{minipage}{0.31\textwidth}
\begin{lstlisting}[caption=HaiQ Constraint Example,label=haiq:constraintexample, language=haiq, breaklines=true, basicstyle=\scriptsize\ttfamily]
all pu: PU | #pu.services > 0
\end{lstlisting}
\end{minipage}
\hfill
\begin{minipage}{0.65\textwidth}
\begin{lstlisting}[caption=HaiQ Properties Example,label=haiq:propertiesexample, language=haiq, breaklines=true, basicstyle=\scriptsize\ttfamily]
label success [some Deployment:finishedOK=true]
property rangeP [F success] as reliability;
property rangeR{executiontimeRew} [F done] as execution_time;
\end{lstlisting}
\end{minipage}
\end{figure}

%% file: 4_QCFramework/qcframework.tex
\begin{figure}[H]
    \centering
    \includegraphics[width=0.482\textwidth]{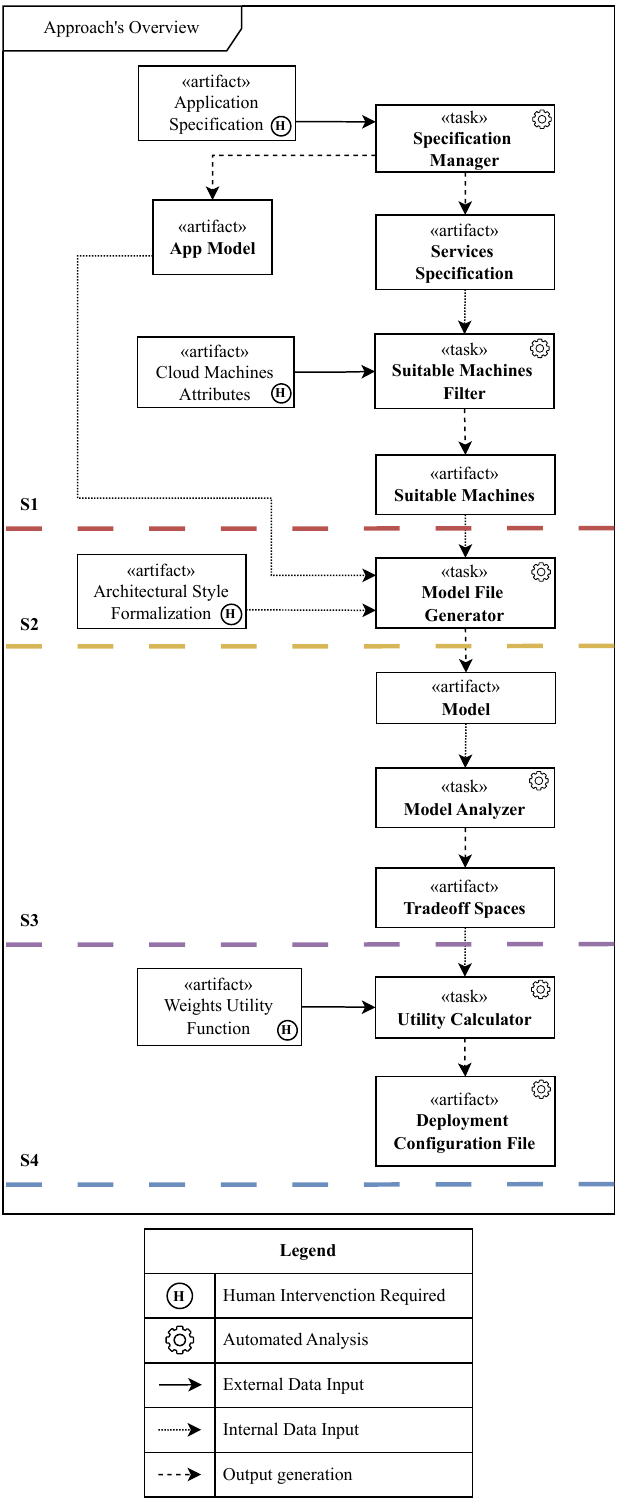}
    \Description{Overview of the design trade-offs analysis system with quantitative guarantees for hybrid service applications}
    \caption{Design Trade-off Space with Quantitative Guarantees Analyzer System for Hybrid Service Applications}
    \label{fig:TOAS}
\end{figure}

The proposed framework enables the analysis of design trade-offs in qualities (e.g., cost, performance, reliability, and quantum error) of quantum-classical hybrid applications (Figure~\ref{fig:TOAS}). 
The process receives as input a specification that contains the sets of requirements and resources available (e.g., machines, services), the behaviour model and constraints of the application, and the architectural style of hybrid service applications. Given this information, the process of generating different configurations and analysing them begins. This procedure is divided into four stages: 
\begin{enumerate}[leftmargin=1.2cm,start=1,label={(\bfseries S\arabic*):}]
    \item {\em Filtering.} In which the elements that are not suitable to build feasible system configurations (e.g., QPUs that do not meet quantum service requirements) are discarded.
    \item {\em Configuration Generation}. During which alternative system configurations are generated. The process receives as input a specification of the available resources (services, suitable machines) and the intended application behavior (described as an abstract workflow in HaiQ), as well as a formalized description of our hybrid quantum-classical architectural style (as a set of first-order predicate logic constraints, cf. Section~\ref{sec:style}). It produces a set of probabilistic models, each of which captures the behavior of a legal architectural configuration.
    \item {\em Configuration Analysis.} The collection of models for each configuration is analyzed using quantitative verification techniques that check system properties formalized in the logic M-PCTL described in Section~\ref{sec:background}.
    \item {\em Configuration Ranking and Selection}. The set of configurations can be ranked and selected according to their qualities. Although multiple criteria can be employed, the current version of the proposal uses utility theory to rank configurations according to different contexts of use that consider concerns such as cost, execution time, reliability, and quantum error.
\end{enumerate}

In the remainder of this section, a description of the formalization of architectural style for service-based quantum-classical applications is presented, followed by a description of each of the stages in the process.

\subsection{Architectural Style}
\label{sec:style}
The possible structures of a family of systems that are related by shared structural and semantic properties are characterized employing an~\textit{architectural style}~\cite{DBLP:books/daglib/0084284}.  
\begin{definition}[Architectural Style]
The architectural style of a quantum-classical service-based application is characterized as a 2-tuple $\mathcal{A} = \langle \Sigma,\mathcal{C}\rangle$, where:
\begin{itemize}
    \item $\Sigma = \langle M, S, K, \Pi, \Lambda \rangle$ is an architectural signature, such that:
    \begin{itemize}
    \item $M = M_q \cup M_c$ is a set of processing unit types (machine types, for short), where $M_q$ and $M_c$ are disjoint sets of quantum processing units (QPU) and classical processing units (CPU) types, respectively. 
    \item $S = S_q \cup S_c$ is a set of service types, where $S_q$ and $S_c$ are disjoint sets of quantum and classical service types, respectively. 
    \item $K$ is a set of connector types.
    \item $M$ is a set of machine types.
    \item $\Pi : (M \cup S \cup K) \rightarrow 2^{\mathcal{P}}$ is a function that assigns sets of symbols typed by datatypes in a fixed set $\mathcal{D}$ to architectural types $\kappa \subseteq M \cup S \cup K$. $\Pi(\kappa)$ represents the properties associated with type $\kappa$ (we denote a property $p \in \Pi(\kappa)$ simply as $\kappa.p$). 
        \begin{itemize}
            \item For machines, $m \in M$, the set of properties $\Pi(m)$ is a tuple $(s, m_{c_r}, m_{c_b}, m_{c_c}, m_{q_q{_{min}}},\\ m_{q_q{_{max}}}, m_{q_s{_{min}}}, m_{q_s{_{max}}}) \in S \times \mathbb{R}^3_{\geq0} $ in which $m.s \in 2^{S}$ corresponds to the set of service types  that can be deployed in $m$, whereas the remaining three propoerties are the memory, bandwidth, and processing capacity of $m$.
            \item For service types $s \in S$ the set of properties $\Pi(s)$ are a tuple $(m, h, s_c, s_r, s_b, s_f, s_a, s_e, s_{s}, s_{d}, l) \in 2^M \times (S_c \cup \emptyset) \times \mathbb{R}^3_{\geq0}$, where $s.m$ corresponds to the set of machine types in which $s$ can be deployed, $s.h$ is a classical service that wraps $s$ if it is a quantum service, and $s.s_c$, $s.s_r$, $s.s_b$, $s.s_f$, $s.s_a$, $s.s_e$, $s.s_s$, $s.s_d$ are the processing capacity minimum requirements, memory, bandwidth, percentage of reliability, online availability, baseline execution time, number of shots and the depth of the quantum circuit. Additionally, $l \subseteq S$ is the set of services that $s$ can communicate with (either invoke or be invoked by).
        \end{itemize}
    \item $\Lambda : (S \cup K) \rightarrow 2^{\mathcal{P}} \cup 2^{\mathcal{R}}$ is a function that assigns a set of symbols typed by a fixed set $\mathcal{R}$ to connectors $\kappa \in K$. $\Lambda(\kappa)$ represents the ports of a service (conversely, roles if $\kappa$ is a connector), which define logical points of interaction with $\kappa$'s environment. 
    To denote a port/role $q \in \Lambda(\kappa)$, we write $\kappa::q$.
   \end{itemize} 
    \item $\mathcal{C}$ is a set of structural constraints that prescribes the set of legal configurations in $\mathcal{A}$. Constraints in $\mathcal{C}$ are expressed in a constraint language based on first-order predicate logic in the style of Alloy~\cite{DBLP:journals/cacm/Jackson19}, analogous to OCL~\cite{Warmer:2003:OCL:861416} and Acme~\cite{DBLP:conf/cascon/GarlanMW97} (e.g., {\sf \small $\forall q:S_q \bullet q.h \in S_c$ } --  ``every quantum service must have a classical service wrapper'').
    \textbf{Table}~\ref{CONSTRAINTS_TABLE} provides an excerpt of several constraints representative of our architectural style.~\footnote{Due to space constraints, we do not reproduce the whole set of constraints included in our formalization. The entire specification, can be downloaded from~\href{https://zenodo.org/records/18659243?preview=1&token=eyJhbGciOiJIUzUxMiJ9.eyJpZCI6IjY4YjI4M2Q2LWE2ZGItNGM0Ni04MTY3LTczMjFkNjM2YTU2YyIsImRhdGEiOnt9LCJyYW5kb20iOiIwMGY3Y2M4ZDVhYWRmNGRiOGZiMTk2YmI0NmViMjkxMyJ9.0yslZv8kT1DgxLxCv9MwL6TSOrfSP2P7cye_3yjDNMwa12uf4m7IFQ746IW4hZF-dBA8PNBlnpZkOx8VxpB08w}{Architectural Style link}} 
\end{itemize}
\end{definition}
\begin{definition}[Configuration] A configuration in an architectural style $\mathcal{A}=\langle \Sigma, \mathcal{C} \rangle$, given a fixed universe of architectural elements $\mathcal{A}_{\Sigma}$, is a graph $\mathcal{G} = (\mathcal{N}, \mathcal{E})$ satisfying the constraints imposed by $\mathcal{C}$, where $\mathcal{N}$ is a set of nodes, such that $\mathcal{N} \subseteq \mathcal{A}_{\Sigma}$, and $\mathcal{E}$ is a set of of pairs typed by $\mathcal{P} \times \mathcal{R}$ that represent attachments between ports in services, and roles in connectors. We denote the type of an architecture element $a \in \mathcal{A}_{\Sigma}$ as $type(a)$.
\end{definition}
\begin{table*}[htbp]
\caption{Architectural Style - Constraints}\label{CONSTRAINTS_TABLE}
\begin{center}
{\scriptsize
\begin{tabular}{|m{3cm}|m{4cm}|m{5cm}|}
\hline
\textbf{Description}& \textbf{FOPL Formalization} & \textbf{Alloy Formalization} \\
\hline
Every quantum service has to be associated with a classical service wrapper.  & $\forall q_s:S_{q} \bullet q_s.h \in S_{c}$ & 
\begin{verbatim}
all qs: QuantumService | 
qs.hybrid_service in ClassicalService
\end{verbatim}
\\
\hline
A quantum service cannot share the classical service with which it is associated.  & $\forall q_s, q_s' : S_{q} \bullet q_s \neq q_s' \Rightarrow q_s.h \neq q_s'.h$ & 
\begin{verbatim}
all qs, qs': QuantumService | 
qs != qs' implies 
qs.hybrid_service != qs'.hybrid_service
\end{verbatim}
\\
\hline
Two services that form a hybrid service are communicated through a link.
  & $\forall s, s' : S \bullet  (s \neq s' \wedge s.h = s') \Rightarrow (s\in s'.l \wedge s' \in s.l )$ & 
\begin{verbatim}
all s, s': Service | 
s !=s' and s' in s.hybrid_service  
implies (s in s'.link and s' in s.link)
\end{verbatim}
\\
\hline
If a service is deployed on a machine, then that machine can only be related to that service.
    & $\forall m\!:\!M, s\!:\!S \bullet s \in m.s \Rightarrow s.m = m$ & 
\begin{verbatim}
all m:PU, s:Service | 
s in p.services implies s.machines = m
\end{verbatim}
\\
\hline
\end{tabular}
}
\end{center}
\end{table*}

\subsection{Filtering (S1)}
\label{sec:filtering}
The process receives as input a specification of the application that consists of two parts. The first part is a formal structural and behavioural description of the application's services, including their hardware and non-functional requirements (cf. \emph{\textbf{Table}~\ref{tab:requirements}}). 
The second part is a specification of the set of available processing units and their attributes, so that they can be adequately matched to suitable services.
\begin{table}[H]
\caption{Non-Functional Service Requirement Attributes}
\begin{center}
\begin{tabular}{|c|p{8cm}|c|}
\hline
\textbf{Attribute}& \textbf{Description} & \textbf{Mandatory} \\
\hline
id & Identifier string of the service & \checkmark \\
\hline
api &  URI for the service's specification. & \checkmark \\
\hline
mode & Indicates the computing model used by the service (quantum or classical). & \checkmark\\
\hline
memory & Variable describing the required memory capacity, in MB. & \checkmark \\
\hline
cpu & Number of required CPU cycles for the service execution. & \checkmark\\
\hline
execution\_time & Maximum execution time (seconds) of the classical service task. & \checkmark \\
\hline
qubits & Number of quantum bits used in quantum algorithm implementation. & $\times^*$ \\
\hline
num\_shots & Number of times the quantum algorithm is executed. & $\times^*$ \\
\hline
number\_request & Number of requests per minute received by the service. & \checkmark \\
\hline
max\_request\_size & Request maximum size in bytes. & \checkmark\\
\hline
availability & Number of hours per day the service is available. & \checkmark\\
\hline
reliability & Percentage representing the probability of successful service execution &\checkmark\\
\hline
instances & Number of service instances. & \checkmark\\
\hline
mandatory & Indicates if the service must appear in the deployment. & \checkmark\\
\hline
\multicolumn{3}{l}{$^{\mathrm{*}}$Attributes presented only in quantum services.}
\end{tabular}
\end{center}
\label{tab:requirements}
\end{table}
According to the service requirements specifications and the hardware characteristics of the cloud providers' machines, a set of suitable machines for each service is obtained. In order to make an assignment between a service and a set of candidate machines, a filter must be applied.
Given the two types of services and processing units (quantum and classical), there is a filter for each of the types:

\begin{enumerate}
    \item Given a classical machine $m{_c} \in M_{c}$ with~\(\ m_{c_r}, m_{c_b}, m_{c_c}\) as the machine's memory, bandwidth and processing capacity and being~\(\ s_{r}, s_{b}, s_{c}\) as the service's minimum memory, bandwidth and processing capacity required, the following restrictions must be accomplished:
    \begin{quote}
    $M^s_c = \{ m \in \; | \; type(m) \in M_c \wedge s.s_r \leq m.m_{c_r} \wedge s.s_b \leq m.m_{c_b} \wedge s.s_c \leq m.m_{c_c}  \}$
    \end{quote}
    \item Given a quantum machine $m{_q} \in M_{q}$ with~\( m_{q_q{_{min}}}, \newline~m_{q_q{_{max}}}, m_{q_s{_{min}}}, m_{q_s{_{max}}}\) as the machine's minimum and maximum number of qubits and minimum and maximum number of shots and being~\(\ s_{q}, s_{s}\) as the service's minimum qubits and shots required, the following restrictions must be accomplished: 
    \begin{quote}
    $M^s_q = \{ m \in \; | \; type(m) \in M_q~\wedge~m.m_{q_q{_{min}}} \leq s.s_q \leq m.m_{q_q{_{max}}}~\wedge~m.m_{q_s{_{min}}} \leq s.s_s \leq m.m_{q_s{_{max}}}\}$

    \end{quote}
\end{enumerate}

As a result, a set of pairs of (services, ({classical machines}, {quantum machines})) is obtained from the services' application and a set of quantum and classical machines $M$. The set of machines that satisfy the constraints for service $s$ is $ M^s \equiv M^s_c \cup M^s_q$, taking into account the two definitions proposed above.

\subsection{Configuration Generation (S2)}
\label{sec:generation}

This stage focuses on the synthesis of the set of potential architectural configurations for the application. 
To this end, both a formalization of the architectural style described earlier in this section, as well as a description of the set of services and the set of suitable machines for each application's service obtained in the previous stage, are integrated into a single formal specification. Such specification is encoded in a subset of the language \textbf{Alloy}, which is employed to define the structural part of a HaiQ model. This process includes the definition of:

\begin{itemize}
    \item \textbf{Signatures:} The creation of structures representing the services and their potential associated machines. In \textit{Listings}~\ref{sig:Service} and~\ref{sig:PU} an example of the two principal signatures is provided:

\begin{figure}[ht]
\begin{minipage}{0.48\textwidth}
\Description{Representation in HaiQ syntax of Service and Processing Unit signatures}
\begin{lstlisting}[caption=Service Signature,label=sig:Service, language=haiq, breaklines=true, basicstyle=\scriptsize\ttfamily]
        abstract sig Service {
        	   machines : some PU,
        	   deployment : one Deployment,
        	   hybrid_service: set Service,
        	   link: some Service
            }
\end{lstlisting}
\end{minipage}
\hfill
\begin{minipage}{0.48\textwidth}
\begin{lstlisting}[caption=Processing Unit Signature,label=sig:PU, language=haiq, breaklines=true, basicstyle=\scriptsize\ttfamily]
    abstract sig PU {
        services : some Service
       }
\end{lstlisting}
\end{minipage}
\end{figure}

\begin{lstlisting}[caption=HSA Signature Examples,label=hsa:signatures, language=haiq, breaklines=true, basicstyle=\scriptsize\ttfamily]
                abstract sig QuantumGroveralg extends QuantumService{}
                abstract sig Binarysearch extends ClassicalService {}
\end{lstlisting}

    The \textit{PU} signature includes the \textit{services} attribute of type \textit{Service}. This attribute denotes the service that is \textit{deployed} on the PU.
    The \textit{Service} signature is defined by the following attributes:
    \begin{itemize}
        \item \textit{machines}: Refers to the machine on which the service is \textit{deployed}. It corresponds to an element $m$ of the set $M^s$ resulting from stage S1.
        \item \textit{deployment}: Refers to the deployment configuration to which the service belongs.
        \item \textit{hybrid\_service}: Corresponds to property $h$ of the architectural style.
        \item \textit{link}: Corresponds to property $l$ described in the architectural style.
    \end{itemize}
    Two examples of signatures from the motivating scenario are showed in Listing~\ref{hsa:signatures}.

    \item \textbf{Constraints:} The definition of association restrictions between services and machines. These limitations are formalized through an \texttt{Alloy} predicate, which defines the model's constraints. Some examples of constraints are collected in the Table~\ref{CONSTRAINTS_TABLE}. Also, specific constraints related to the hybrid application are used to build de HaiQ model. \textit{Listing}~\ref{const:HSA} collects two examples of the motivating scenario.
    \begin{lstlisting}[caption=HSA Constraint Examples,label=const:HSA, language=haiq, breaklines=true, basicstyle=\scriptsize\ttfamily]
all ag:Aggregator | #(ag.link & (QuantumGroveralg + Resultprocessing + Aggregator))= 0 and #(ag.link & (Binarysearch + Groveralg)) > 0
all qg: QuantumGroveralg | #(qg.link & (Sensor + Aggregator + Resultprocessing)) = 0
    \end{lstlisting}
    
    \item \textbf{Execution Scope:} For the generation of the set of potential configurations for the application, the minimum number of execution instances for the predicate must be specified, ensuring the existence of at least one machine instance for each service. For example, if after phase S1, a set of three classical machines has been obtained for the Motivating Scenario, and its total number of classical services is six, then for each machine obtained, at least six instances of that machine must be generated, one for each service. This ensures that there are enough instances for all services. This is specified in the HaiQ model with the statement included in the Listing~\ref{ee:HSA}.
    \begin{lstlisting}[caption=Execution Escope,label=ee:HSA, language=haiq, breaklines=true, basicstyle=\scriptsize\ttfamily]
                                  run show for 6
    \end{lstlisting}
    
\end{itemize}

For the generation of the configuration space we employ the HaiQ analyzer tool, which is able to synthesize valid architectural configurations satisfying the defined constraints. The resulting set of configurations enables the analysis of properties of interest across the solution space, which is described in the next section.

\subsection{Configuration Analysis (S3)}
\label{sec:analysis}

This stage concerns the quantitative analysis of trade-offs across the solution space generated in the previous phase. To conduct this analysis, the model is enriched with detailed information regarding behavioral and quantitative aspects (e.g., quality attributes encoded as costs or rewards) of various architectural elements, such as those that represent services and machines.

The analysis of the configuration space is carried out by evaluating the four key Quality of Service (QoS) dimensions established in the Table~\ref{tab:qos} (\textit{cost}, \textit{execution time}, \textit{reliability}, and \textit{quantum error}). The calculation of each dimension is implemented via a reward system, where a value is assigned to each state transition action (e.g., the pairing between a service and a machine). This process yields the quantitative value associated with each dimension for every configuration.

The methodology for obtaining the metrics for each studied dimension is detailed below, illustrated with examples from the motivating scenario:

\begin{itemize}
    \item \textbf{Cost}: Represents the total economic cost associated with using the configuration. It is calculated as the weighted sum of the usage costs of both classical and quantum machines. The reward for each service is calculated based on its nature (classical~\ref{pro:Cost_Classical} or quantum~\ref{pro:Cost_Quantum}) and is typically measured in terms of \textit{price/request}. \textit{Listing}~\ref{pro:Cost_Classical} and \textit{Listing}~\ref{pro:Cost_Quantum} detail the reward for each type.

\begin{lstlisting}[caption=Classical Service - Cost Reward,label=pro:Cost_Classical, language=haiq, breaklines=true, basicstyle=\scriptsize\ttfamily]
//Classical service
reward costRew [machines:cpair] true : machines.cpucostfactor/(numberrequest*60);
\end{lstlisting}
\paragraph{}
\begin{lstlisting}[caption=Quantum Service - Cost Reward,label=pro:Cost_Quantum, language=haiq, breaklines=true,  basicstyle=\scriptsize\ttfamily]
// Quantum Service
reward costRew [machines:qpair] true : ((machines.qpuprize)/60)*(depth/machines.clops)*shots;
\end{lstlisting}
    In the case of a hybrid service, the total cost is the result of the addition of the cost of its classic and quantum services.
    \item \textbf{Execution Time}: This metric, quantified in \textit{nanoseconds/request}, reflects the time required to complete the task associated with the service. The execution time is directly dependent on the characteristics of the associated machine. A machine with higher capabilities executes the corresponding service task more rapidly.
\begin{lstlisting}[caption=Classical Service - Execution Time Rewards,label=pro:ExecutionTime_Classical, language=haiq, breaklines=true, basicstyle=\scriptsize\ttfamily]
// Classical Service
reward executiontimeRew [machines:cpair] true : msexecutiontime*(mslogicalperformancefactor*msram*msbandwidth)/(machines.cpulogicalperformancefactor*machines.cpuram*machines.cpubandwidth);
reward executiontimeRew [machines:cpupenalty] true : msexecutiontime*(mslogicalperformancefactor*msram*msbandwidth)/(machines.cpulogicalperformancefactor*machines.cpuram*machines.cpubandwidth);
reward executiontimeRew [machines:mspenalty] true : msexecutiontime*(mslogicalperformancefactor*msram*msbandwidth)/(machines.cpulogicalperformancefactor*machines.cpuram*machines.cpubandwidth);
\end{lstlisting}
\begin{lstlisting}[caption=Quantum Service - Execution Time Reward,label=pro:ExecutionTime_Quantum, language=haiq, breaklines=true, basicstyle=\scriptsize\ttfamily]
// Quantum Service
reward executiontimeRew [machines:qpair] true : (depth/machines.clops)*shots;
\end{lstlisting}
    The execution time is quite different between the two existing types of services. In the classic ones (detailed in \textit{Listing}~\ref{pro:ExecutionTime_Classical}), the total execution time includes the time of executing the request and the penalty times. In this scenario, the two penalty times \texttt{[machines:cpupenalty]} and \texttt{[machines:mspenalty]} correspond to hardware and software failures. These penalties have been established as the total request execution time. On the other hand, the quantum service only takes into account the execution time of the quantum algorithm, as it is illustrated in \textit{Listing}~\ref{pro:ExecutionTime_Quantum}.

    \item \textbf{Reliability}: Service reliability is a composite factor that includes both a \textit{hardware} and a \textit{software} component. Reliability is only associated with classical services and during the service-machine pairing phase (service association/deployment on a machine, see in \textit{Listing}~\ref{pro:Reliability_Hardware})  is determined by the hardware component. In contrast, reliability during the operational phase (upon receiving the process request) is related to the stability of the software component as it is shown in \textit{Listing}~\ref{pro:Reliability_Software}.
\begin{lstlisting}[caption=Hardware Reliability Transition,label=pro:Reliability_Hardware, language=haiq, breaklines=true, basicstyle=\scriptsize\ttfamily]
// Hardware Transition
[machines:cpair] (currentstatus=initial) -> machines.cpuerrorrate: (currentstatus'=cpuerror) & (penalty'=true) + 1-machines.cpuerrorrate: (currentstatus'=deployed);
\end{lstlisting}

\begin{lstlisting}[caption=Software Reliability Transition,label=pro:Reliability_Software, language=haiq, breaklines=true, basicstyle=\scriptsize\ttfamily]
// Software Transition
[deployment:SearchActivationCall] (currentstatus=ready) -> mserrorrate: (currentstatus'=mserror) & (penalty'=true) + 1-mserrorrate:(currentstatus'=running);
\end{lstlisting}

    \item \textbf{Quantum Error}: This metric quantifies the number of executions that provide wrong results obtained by the service. Given the nature of the current NISQ (\textit{Noisy Intermediate-Scale Quantum}) era of quantum computers, these devices introduce significant noise into execution results, making this a critical factor to consider. Its corresponding reward is detailed in \textit{Listing}~\ref{pro:QuantumError_Quantum}. For classical services, the quantum error is considered \textit{null} as it is reflected in \textit{Listing}~\ref{pro:QuantumError_Classical}.

\begin{lstlisting}[caption=Classical Service - Quantum Error Reward,label=pro:QuantumError_Classical, language=haiq, breaklines=true, basicstyle=\scriptsize\ttfamily]
// Classical service
reward quantumerror [machines:cpair] true : 0;
\end{lstlisting}
\begin{lstlisting}[caption=Quantum Service - Quantum Error Reward,label=pro:QuantumError_Quantum, language=haiq, breaklines=true, basicstyle=\scriptsize\ttfamily]
// Quantum service
reward quantumerror [machines:qpair] true : (((machines.readoutminerror)+(machines.readoutmaxerror))/2)*shots;
\end{lstlisting}
\end{itemize}

Considering these metrics, the probabilistic model checking tool PRISM, integrated into the HaiQ environment, evaluates each of the defined properties across the entirety of the trade-off space. This process yields a set of valid configurations, each associated with a set of quantitative values for the studied quality dimensions.

\subsection{Configuration Ranking and Selection (S4)}
\label{sec:ranking}

This last stage of the process involves the evaluation of the various configurations with the objective of obtaining the optimal solution of the assessed model. This part of the process incorporates user preferences regarding the relative importance of the various dimensions of concern (e.g., timeliness, quantum error). The result of this phase will be the solution(s) that better optimize such preferences. Although there are various ways to identify the solutions that best match user preferences, in this case, we use Utility Theory, which enables us to seek the outcome that maximizes user satisfaction. To accomplish this, we employ a simple weighted additive utility function:

$$
U(x) = \sum_{i=1}^{n} a_i x_i
$$

where:
\begin{itemize}
    \item $U(\mathbf{x})$: The total utility value for a set of dimensions $\mathbf{x} = (x_1, x_2, \ldots, x_n)$.
    \item $n$: The total number of dimensions or features considered in the model.
    \item $x_i$: The value of the $i$-th dimension or characteristic being studied. In this case, this value could represent execution time, cost, reliability, and quantum error.
    \item $a_i$: The weight or coefficient assigned to the $i$-th dimension. These weights are typically normalized such that their sum is equal to 1.
\end{itemize}

The utility-weighted function used in the \textit{Evaluation Section~\ref{sec:evaluation}} is:

$$
U(x) = w_{cost}~x_{cost} + w_{executiontime}~y_{performance}
$$
$$ + w_{reliability}~z_{reliability} + w_{quantumerror}~w_{quantumerror}
$$
\\
where $w_{cost}, w_{executiontime}, w_{reliability}, w_{quantumerror}$ are coefficients defined by the user for their corresponding dimension, such that ($w_{cost}+w_{performance}+w_{reliability}+w_{quantumerror}=1$).

Finally, obtaining the set of best configurations $\left( \mathbf{x}^* \right)$ is the result of applying:

$$
\mathbf{x}^* = \operatorname{argmax}_{\mathbf{x}} \left( f(x)\right)
$$

In order to apply the utility function, the values of the analyzed dimensions were normalized to a common [0, 1] range. The goal of this transformation is for 0 to represent the minimum utility and 1 to represent the maximum utility. Since the results of the analysis contain metrics that need to be maximized (like \textit{reliability}) and minimized (like \textit{cost}, \textit{execution time}, and \textit{quantum error}), the Min-Max normalization method was adjusted based on each variable's nature, inverting the scale where required.

%% file: 5_Evaluation/evaluation.tex
The purpose of this evaluation section is to explore the research questions described in the introduction, using as a vehicle to do so a prototype implementation of our approach. This section is structured as follows: firstly, the \textit{Experimental Setup} section~\ref{subsec:experimentalsetup} details the conditions under which the experiments were conducted. Secondly, the Results~\ref{subsec:results} section presents an analysis of the experiments. Finally, the \textit{Discussion} section~\ref{subsec:discussion} provides the answers to the proposed research questions (\textbf{RQ1} and \textbf{RQ2}).

\subsection{Experimental Setup}
\label{subsec:experimentalsetup}
To evaluate our approach, two use cases were selected. The first is the Hybrid Search Application, previously detailed in Section \ref{sec:motivating}. The second use case is a Weather Forecast Hybrid System detailed in Section~\ref{subsec:wfhs}.

To run our experiments, the following computer and software requirements have been used:
\begin{itemize}
    \item macOS Ventura (version 13.7.1)
    \item Intel Core i5 with dual-core, running at 2.3GHz.
    \item 16GB of RAM.
    \item HaiQ Relational Probabilistic Model Analyzer v0.2a
\end{itemize}

The final HaiQ generated models are published in this Zenodo repository~\footnote{\href{https://zenodo.org/records/18658906?preview=1&token=eyJhbGciOiJIUzUxMiJ9.eyJpZCI6ImY0NDRkZDVlLTljOTMtNGYwNy04MDJjLWE4NzZkYjFhMzU0MCIsImRhdGEiOnt9LCJyYW5kb20iOiI5YWM4MjMxMzdmYWQ1NmYxNjM1Y2QxMTg4MmFjNjliYSJ9.ZXrKwI_OxrXPp3ni83XU7XKWmCktDfyiHXSy2BwBVYNi7Ck47fayY3TXr-3UcaZJxRycIocQZuqYxquqcC65XA}{HaiQ Models}}.
To evaluate the generated HaiQ models, four M-PCTL properties targeting reliability, execution time, cost, and quantum error have been defined. These properties leverage cumulative rewards that are computed until the system reaches a termination state (labeled as \textit{success} or \textit{done}).

\subsubsection{Implemented Analyzing Tool}
The analysis tool was developed using \textit{Python 3.11.9}. The workflow initiates with the ingestion of system specifications through a web-based interface. A core feature of the implementation is the automated generation of HaiQ models derived from the developed architectural style of hybrid (quantum-classical) service applications and the application specification. The data distribution across the system’s components is handled via a pub-sub protocol implemented with Apache Kafka. Once the simulation finishes, the architectural configurations based on the applied multi-dimensional utility function are ranked and shown on the website. This quantitative analysis is complemented by a specialized data processing pipeline using Pandas~\footnote{\href{https://pandas.pydata.org/}{Pandas}} and Jupyter Notebooks~\footnote{\href{https://jupyter.org/}{Jupiter Notebook}}, which facilitates the rendering of decision region maps and statistical visualizations.

\subsection{Weather Forecast Hybrid System}

\begin{figure}[ht]
    \centering
    \includegraphics[width=0.7\textwidth]{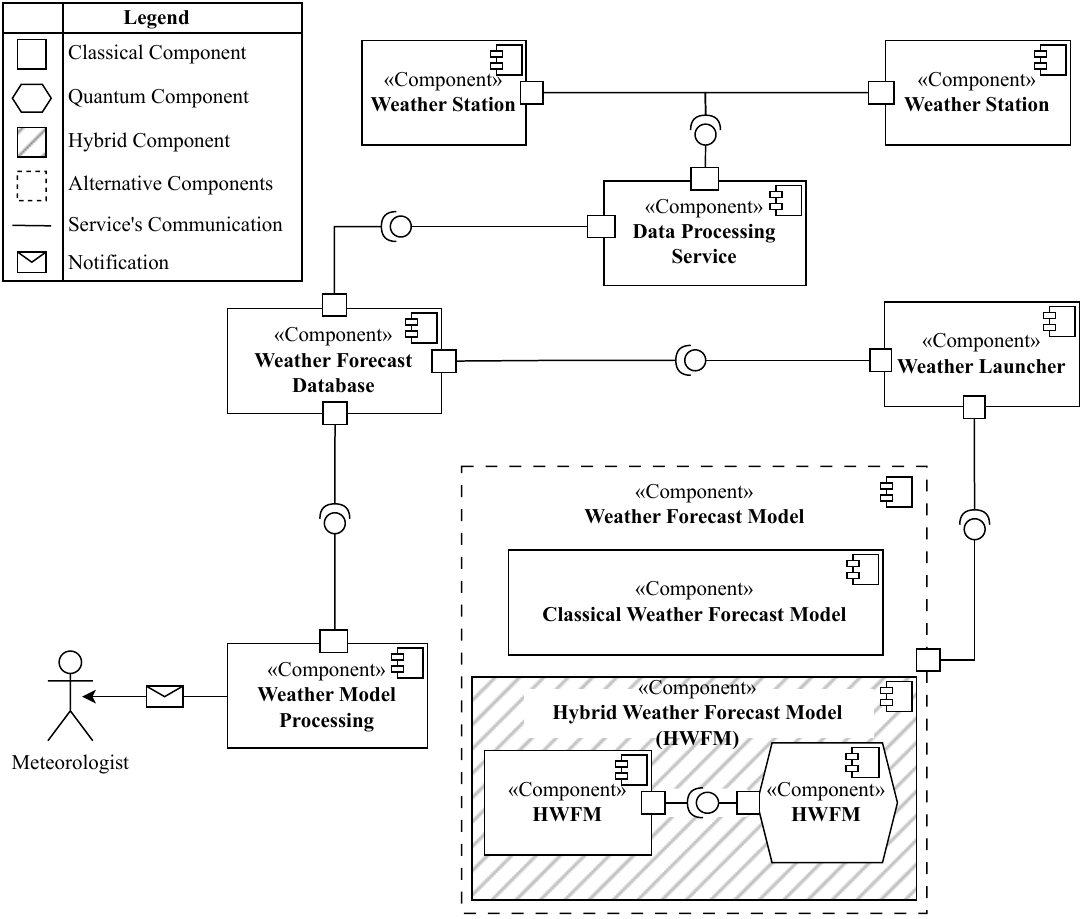}
    \Description{The hybrid weather forecasting system comprises a total of seven services. Two weather station services are connected to the data processing service. This service is connected to a database. The forecast model launch service is connected to that database and is responsible for launching the classic or hybrid forecast model. The weather forecast model processing service reads the data from the database service to display it to the meteorologist.}
    \caption{Architecture of the WFHS System}
    \label{fig:ArchWFHS}
\end{figure}

The architecture depicted in Figure~\ref{fig:ArchWFHS} reflects a simplified structure of the weather forecasting system presented in~\cite{DBLP:conf/wcre/AparicioMoralesHMGBSMM24}. The primary objective of these systems is to provide routine weather forecasts. However, there are exceptional use contexts that require of obtaining rapid predictions, especially in climate emergency situations (such as isolated high-level depressions or DANAs, flash floods, tornadoes, among others) where the window of opportunity to issue a warning is critical for civil safety.

Structurally, these systems typically comprise a data collection layer, represented in our case study by two weather stations (WSS). This information is sent to an intermediate node, the Data Processing Service (DPS), which is responsible for structuring and organizing the data. Following this processing, the information is stored in a database (Weather Forecast Database, WFD) that feeds the forecasting models.

These models are triggered by the Weather Launcher Service (WLS). Once the execution is complete, the resulting information is stored back into the database. Finally, the results processing service (WMP) retrieves the prediction to be interpreted by a meteorologist. In this study, the weather models can be either purely classical (CWFM) or hybrid, incorporating quantum components (HWFM). Their selection will depend on the urgency and context of the forecast.

Notably, this WFHS use case considers alternatives not only for the forecast model (classical vs. quantum) but also for the WSS, DPS, and PRS classical components.
\label{subsec:wfhs}

\subsection{Results}
\label{subsec:results}
After conducting the configuration space exploration for both use cases with the HaiQ tool, a total of \textbf{7680 instances} (1536 classical vs 6144 hybrid instances) were obtained for the motivating scenario (HSA) and \textbf{3456 instances} (1152 classical vs 2304 hybrid instances) for the WFHS.

Table~\ref{tab:haiq_results} shows the total analysis time for each case. It is worth noting that, despite the notable difference in the volume of instances generated ($\Delta = 4224$), the analysis times present great similarity. This circumstance is attributed to the variability in structural complexity and the state space generated by the models due to their behavior. Likewise, the evaluation times for each dimension studied (\textit{cost}, \textit{execution time}, \textit{reliability}, and \textit{quantumerror}) are broken down in said table.
Next, the analysis of the results is presented, starting with the HSA, followed by the HWFS.

\begin{table}[H]
\caption{Evaluation time of each dimension}\label{tab:haiq_results}
\centering
\begin{tabular}{|c|c|c|c|c|c|c|}
\hline
\textbf{\makecell{Use Case}} & 
\textbf{\makecell{Instances\\Generated}} & 
\textbf{\makecell{Evaluation\\Time (h)}} & 
\textbf{\makecell{Cost \\ (s)}} & 
\textbf{\makecell{Execution\\Time (s)}} &  
\textbf{\makecell{Reliability (s)}} &  \textbf{\makecell{Quantum\\Error(s)}}
\\
\hline
HSA & 7680 & 4.18 & 3463.2 & 3481.5 & 4544.3 & 3418.6
\\
\hline
WFHS & 3456 & 4.08 & 3595.4 & 3600.6 & 3820.7 & 3625.5
\\
\hline
\end{tabular}
\end{table}

\subsubsection{Hybrid Search Application}
\begin{figure}[ht]
    \centering
    \includegraphics[width=0.7\linewidth]{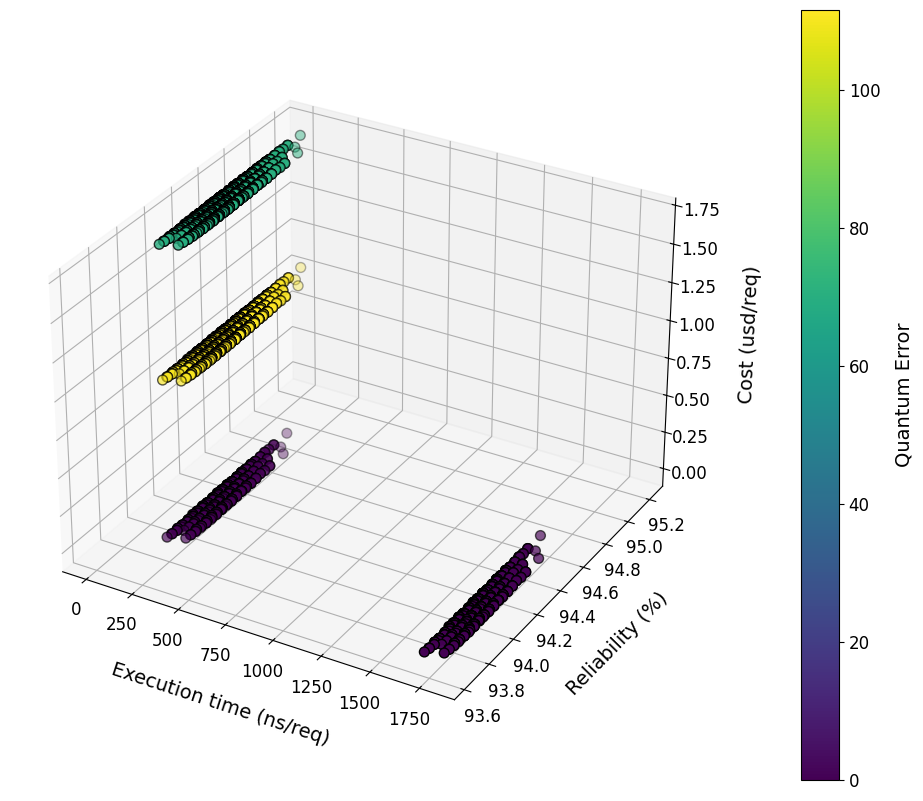}
    \Description{A scatter plot showing four dimensions of the generated configurations, organized into four distinct clusters. The first cluster is characterized by high execution times (between 1500 and 1750 units), high reliability (93.8\% to 95\%), and a cost near zero. The other three clusters share lower execution times (between 100 and 250 units) and similar reliability (93.8\% to 95\%). Among these three, the first sub-cluster has a cost near zero and a quantum error close to zero. The second sub-cluster shows a cost between \$0.75 and \$1.00 per request with a high quantum error of over 100 shots. The third sub-cluster presents a cost between \$1.50 and \$1.75 per request with a medium quantum error of approximately 80 shots.}
    \caption{Design Space of Hybrid Search App}
    \label{fig:hsa}
\end{figure}

As detailed previously in Section~\ref{sec:motivating} (\textit{Motivating Scenario}), the Hybrid Search Application serves as a compact practical example of a hybrid system used to validate the proposed method. This application is composed of four types of services: three sensor services, an aggregator service, a search service (either classical or quantum), and a result processing service. In the Figure~\ref{fig:hsa}, all the possible configurations are represented.

The primary objective of this analysis is to demonstrate the validity of the proposed system in facilitating architectural deployment decisions for systems integrating equifinal services. Specifically, the aim is to explore the entire design space to identify the most suitable service-machine pair configuration. This selection is driven not only by constraints, such as execution time limits and tolerance to quantum error, but also by the pursuit of dynamic adaptation according to potentially changing user preferences.

To evaluate this adaptability, the set of configurations was subjected to five distinct scenarios. Each scenario represents a variation in the weights assigned to each studied dimension within the utility function:

\begin{table}[H]
\caption{Weight configurations for different optimisation scenarios.}\label{tab:hsa_scenarios}
\centering
\begin{tabular}{lcccc}
        \hline
        \textbf{Scenario} & $w_{cost}$ & $w_{executiontime}$ & $w_{reliability}$ & $w_{quantumerror}$ \\
        \hline
        \textbf{1 - Cost priority} & 0.7 & 0.1 & 0.1 & 0.1 \\
        \textbf{2 - Execution time priority} & 0.1 & 0.7 & 0.1 & 0.1 \\
        \textbf{3 - Reliability priority} & 0.1 & 0.1 & 0.7 & 0.1 \\
        \textbf{4 - Quantum error priority} & 0.1 & 0.1 & 0.1 & 0.7 \\
        \textbf{5 - Equal importance} & 0.25 & 0.25 & 0.25 & 0.25 \\
        \hline
    \end{tabular}
\end{table}

\begin{figure*}[t!] 
    \centering
    
    \begin{subfigure}[b]{0.48\textwidth}
        \centering
        \includegraphics[width=\textwidth]{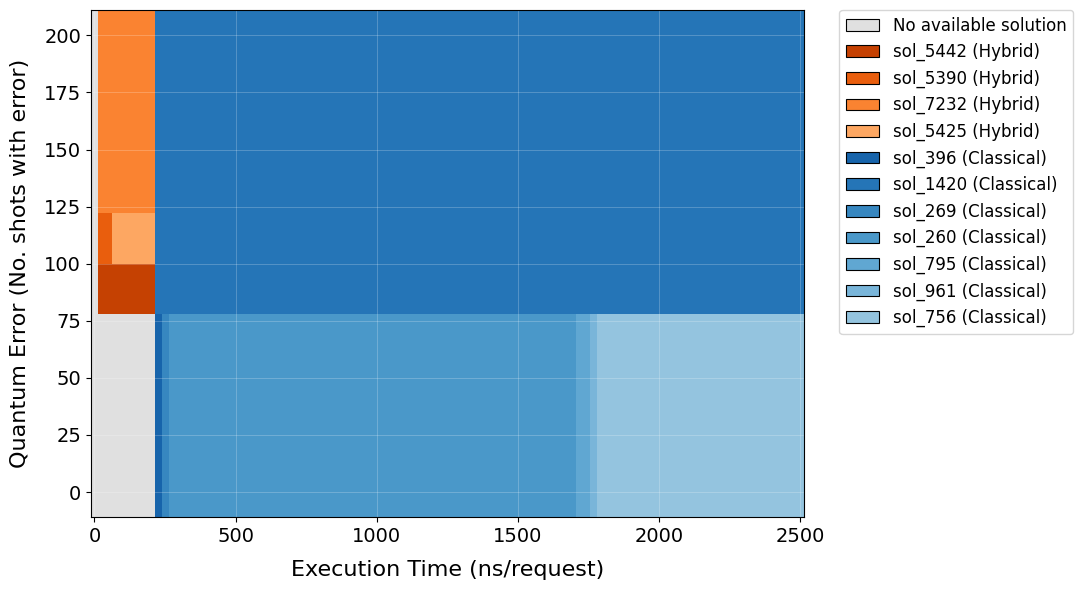}
        \caption{Scenario 1: Cost Priority}
        \label{fig:hsa-s1}
    \end{subfigure}
    \hfill
    \begin{subfigure}[b]{0.48\textwidth}
        \centering
        \includegraphics[width=\textwidth]{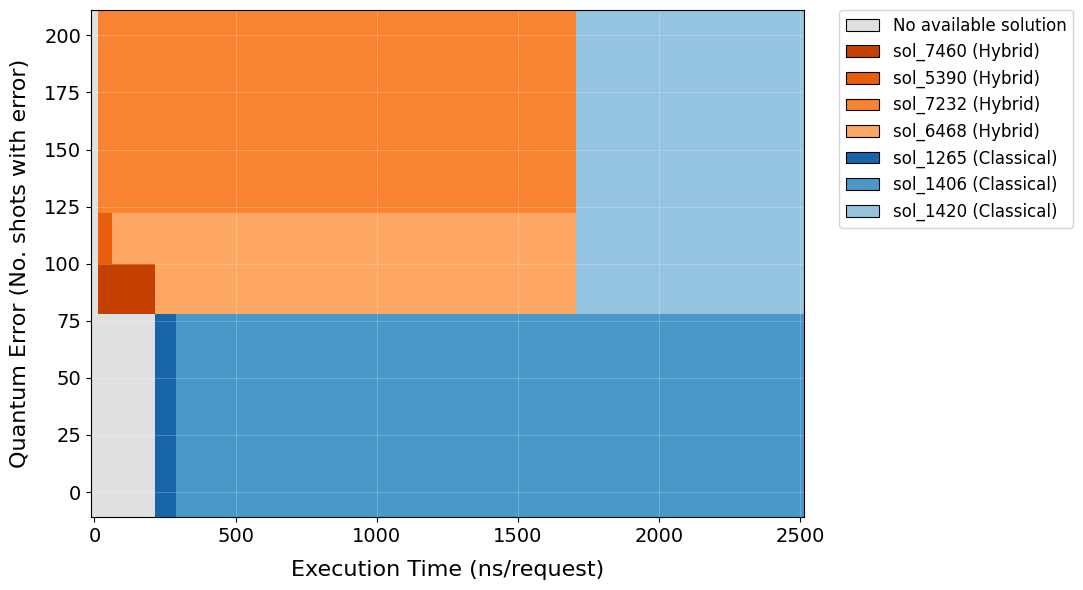}
        \caption{Scenario 2: Time Priority}
        \label{fig:hsa-s2}
    \end{subfigure}
    
    \paragraph{}
    
    \begin{subfigure}[b]{0.48\textwidth}
        \centering
        \includegraphics[width=\textwidth]{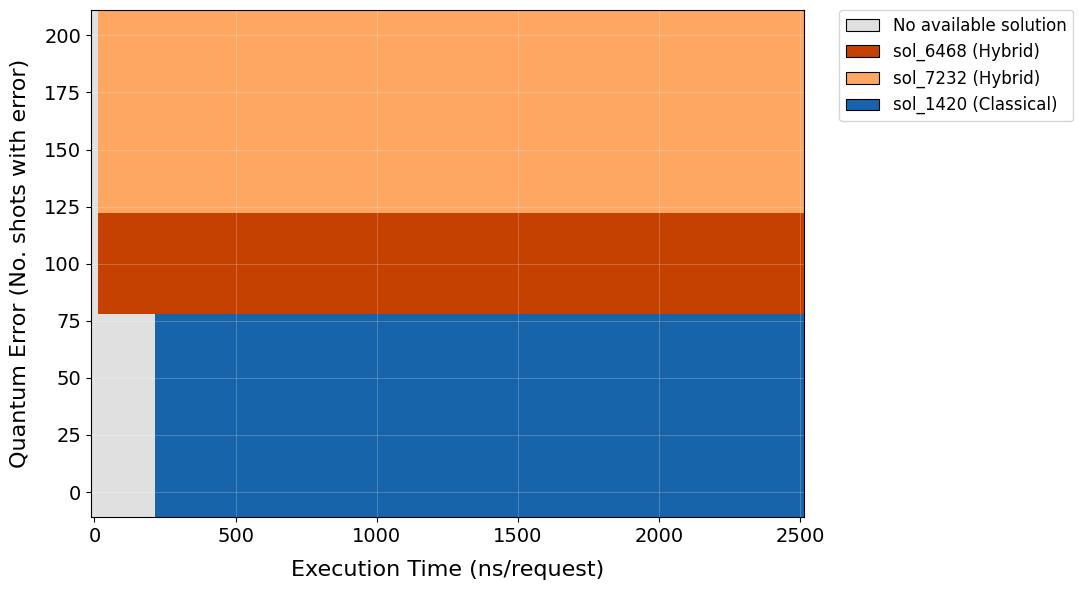}
        \caption{Scenario 3: Reliability Priority}
        \label{fig:hsa-s3}
    \end{subfigure}
    \hfill
    \begin{subfigure}[b]{0.48\textwidth}
        \centering
        \includegraphics[width=\textwidth]{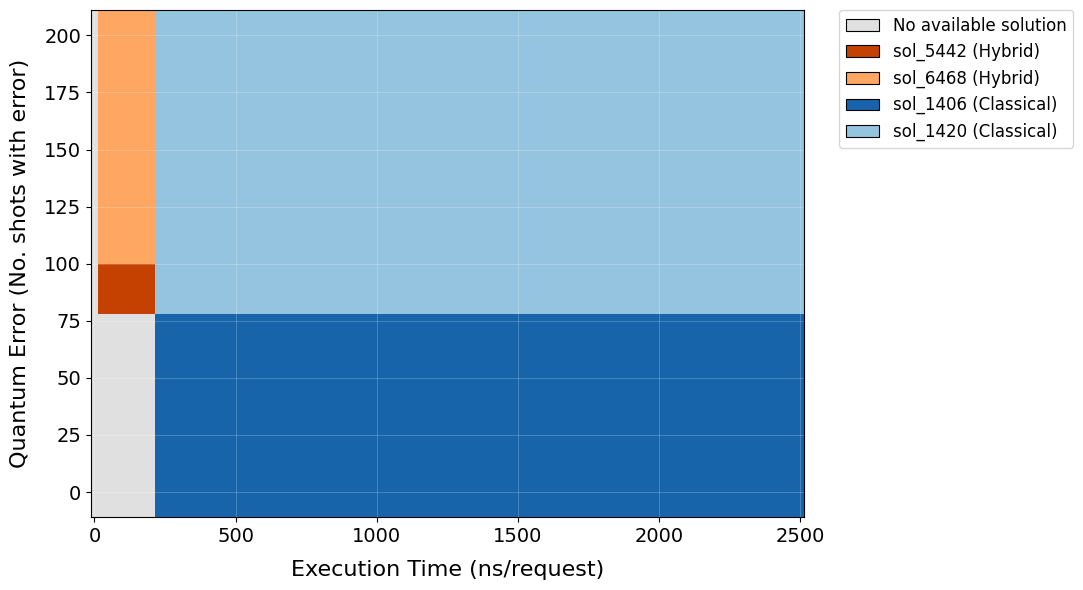}
        \caption{Scenario 4: Quantum Error Priority}
        \label{fig:hsa-s4}
    \end{subfigure}
    
    \paragraph{}
    
    \begin{subfigure}[b]{0.48\textwidth}
        \centering
        \includegraphics[width=\textwidth]{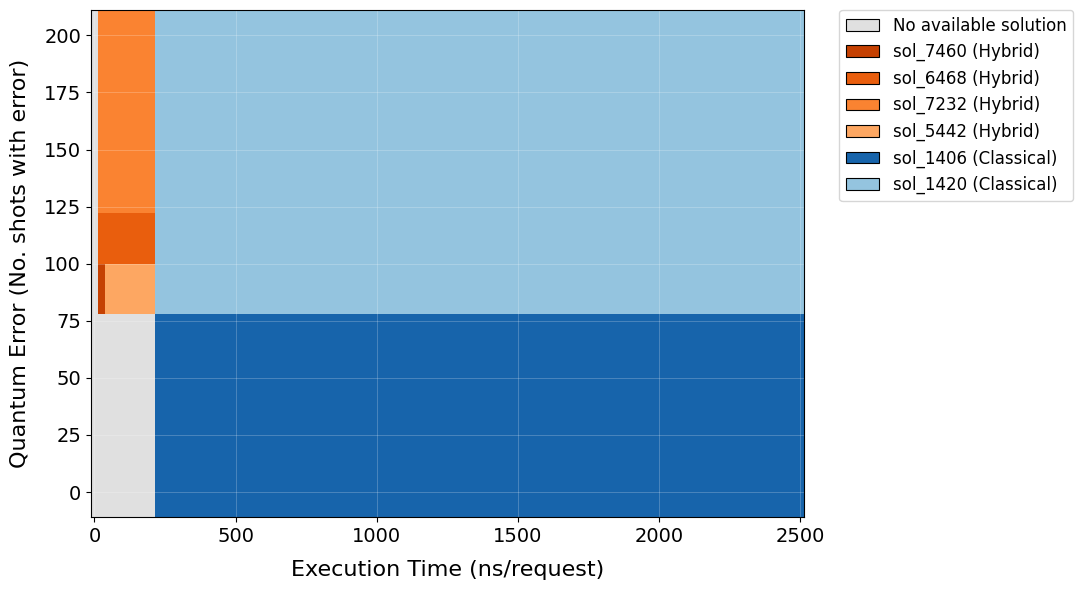}
        \caption{Scenario 5: Balanced}
        \label{fig:hsa-s5}
    \end{subfigure}

    \Description{Distribution of optimal solutions among quantum error and execution time dimension by five different priority scenarios, cost, execution time, reliability, quantum error, and balanced.}
    \caption{Decision region representation of the optimal configurations for the Hybrid Search Application across five different scenarios.}
    \label{fig:hsa-heatmap-scenarios}
\end{figure*}

The results for these scenarios is visualized in \textit{Figures}~\ref{fig:hsa-s1}-\ref{fig:hsa-s5}. These figures display the resulting decision region maps for each scenario, representing the optimal solution type selected under varying constraints of execution time and quantum error. The selection of these two variables is justified as they constitute the primary competing constraints in quantum software engineering. On the one hand, execution time reflects the configuration's performance, while on the other, quantum error captures the fidelity limitations inherent to current NISQ computers. Ultimately, both determine the architectural decision boundary.

Regarding the figures, the following interpretations can be done:
\begin{itemize}
    \item In Scenarios 1 and 4, where cost and result precision are prioritized respectively, a clear dominance of classical configurations over hybrid ones is visualized. Hybrid configurations only begin to be selected when execution time constraints are strict (execution time < 200). In this sense, the proposed system discards the hybrid option due to its high costs and the presence of execution errors stemming from the NISQ era of quantum computers.
    
    \item In Scenarios 2 and 3, the system proposes the hybrid option as the dominant type of configuration across a large portion of the design space. This is because of a smaller cost priority ($w_{cost}$ = 0.1), compared to other scenarios; hybrid solutions are less penalized. This validates the premise that in environments where execution times or reliability must be prioritized over budget, the architecture should be oriented towards configurations that provide results in a timely manner.
    
    \item Thirdly, in the balanced scenario, the system prioritizes classical solutions, maintaining control over cost and precision. In contrast, for stricter time constraints, the set of hybrid configurations becomes optimal as classical solutions cannot operate within those limits.
    
    \item Finally, it is noteworthy that in all scenarios, there exists a set of constraints strict enough to indicate that, given the characteristics of the established services for the use case and the selected set of machines, no valid solutions are yielded for those criteria. In this regard, the proposed system informs the software architect of the existence of a design space where the physical boundaries of the current system are delimited. Thus, the system acts as a diagnostic tool, indicating the possible need to refactor the architecture.
\end{itemize}

\subsubsection{Weather Forecast Hybrid System}
To provide a generalization of the obtained results and mitigate potential domain-specific dependencies, the analysis proceeds to the second proposed case study; the Hybrid Weather Forecast System.

The main objective of this scenario is to evaluate the architectural trade-off between classical solutions, which offer high-precision predictions at the expense of high computational time, and hybrid configurations, which enable faster result generation by sacrificing exactitude. This analysis is conducted within the constraints imposed by the current state of NISQ devices on the design space.

\begin{figure}[ht]
    \centering
    \includegraphics[width=0.7\linewidth]{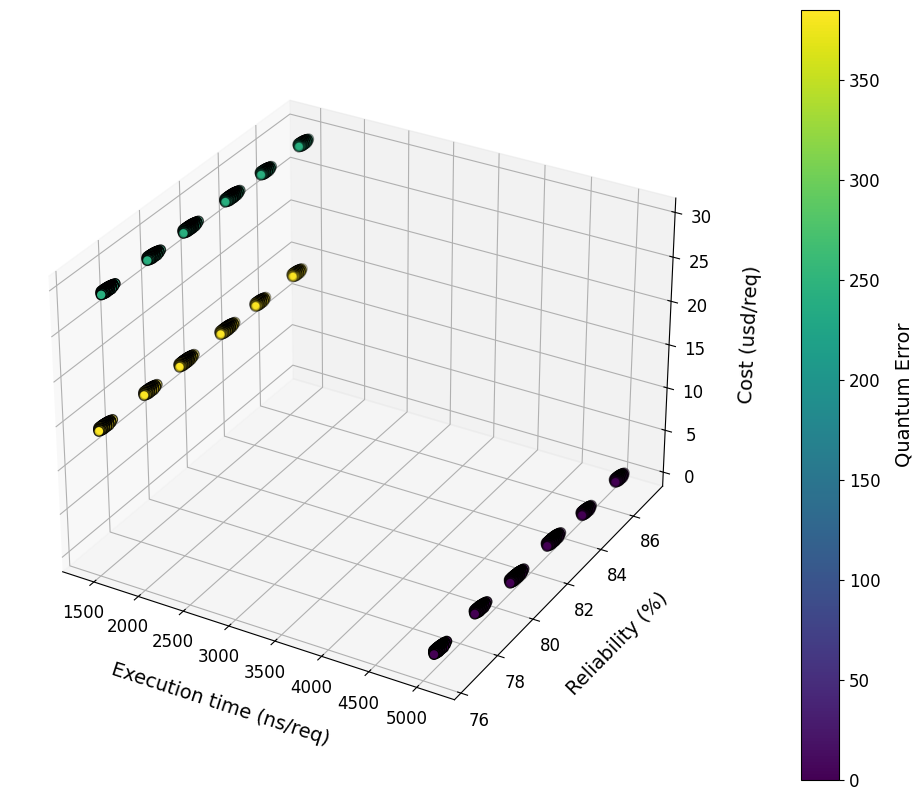}
    \Description{A scatter plot showing four dimensions of the generated configurations, organized into three distinct clusters. The first cluster is characterized by high execution times (between 4800 and 5000 units), a variety of reliability values (77\% to 88\%), and a cost near zero. The other two clusters share lower execution times (between 100 and 1500 units) and similar reliability values. Among these two, the first sub-cluster has a cost near \$15 per request and a quantum error over 350 shots. The second sub-cluster shows a cost near \$30 per request with a lower quantum error of approximately 225 shots.}
    \caption{Solution Space of Weather Forecast Hybrid System}
    \label{fig:wfhs}
\end{figure}

\paragraph{\textbf{5.3.2.1 Global Solution Space Exploration}} \textit{Figure}~\ref{fig:wfhs} shows the generated 3456 configurations. Visual analysis reveals a topology characterized by three completely disjoint data clusters, the separation of which is attributed to the nature of the utilized computational hardware. Specifically, a first cluster is identified corresponding to the classical solutions (1152 configurations), alongside two additional clusters associated with hybrid solutions, differentiated by the QPU provider (IBM and IQM, with 1152 configurations for each one).

\begin{figure}[H]
    \centering
    \begin{subfigure}[b]{0.48\textwidth}
        \centering
        \includegraphics[width=\linewidth]{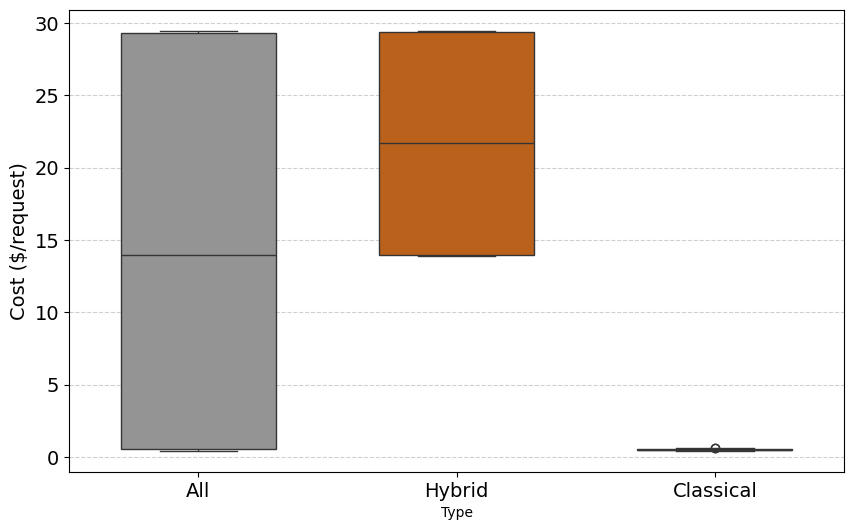}
        \caption{Cost}
        \label{fig:wfhs-cost-comparison}
    \end{subfigure}
    \hfill
    \begin{subfigure}[b]{0.48\textwidth}
        \centering
        \includegraphics[width=\linewidth]{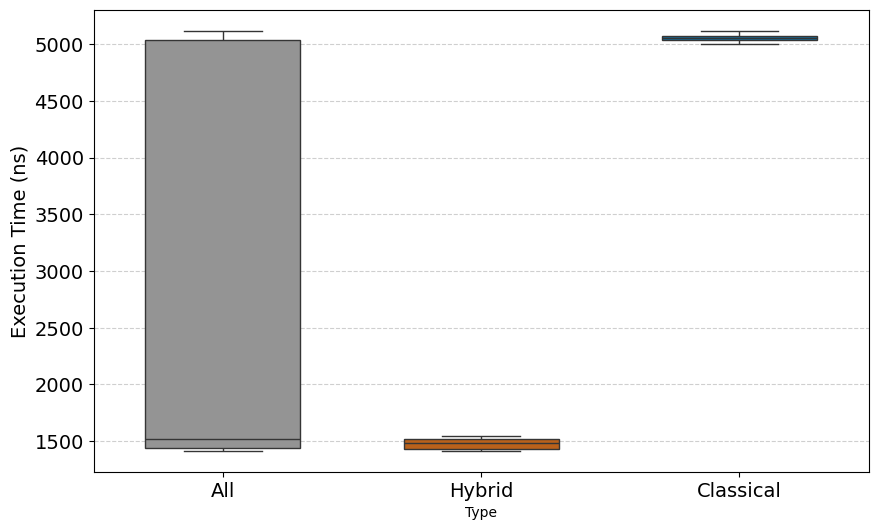}
        \caption{Execution Time}
        \label{fig:wfhs-extime-comparison}
    \end{subfigure}
    
    \paragraph{}
    
    \begin{subfigure}[b]{0.48\textwidth}
        \centering
        \includegraphics[width=\linewidth]{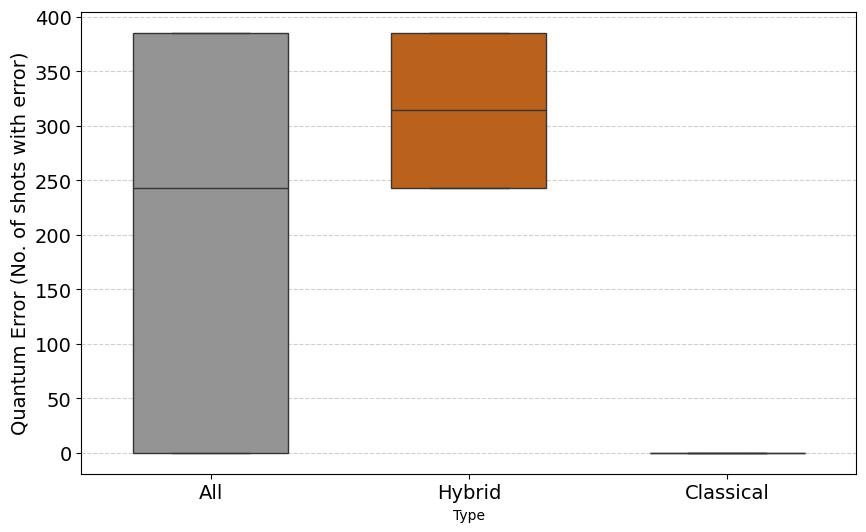}
        \caption{Quantum Error}
        \label{fig:wfhs-quantumerror-comparison}
    \end{subfigure}
    \hfill
    \begin{subfigure}[b]{0.48\textwidth}
        \centering
        \includegraphics[width=\linewidth]{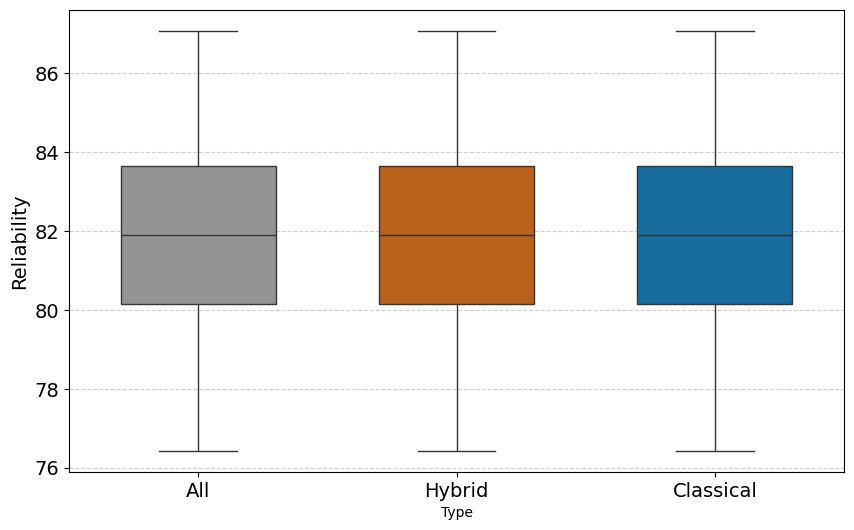}
        \caption{Reliability}
        \label{fig:wfhs-reliability-comparison}
    \end{subfigure}
    
    \Description{Boxplot representation of each dimension, cost, execution time, quantum error, and reliability}
    \caption{Comparison of dimensions by configuration type}
\end{figure}

Upon examining Figure~\ref{fig:wfhs}, which depicts the architectural trade-offs, classical solutions dominate the result quality dimension (ensuring zero quantum error. It is appreciated in \textit{Figure}~\ref{fig:wfhs-quantumerror-comparison}). Conversely, they incur significantly higher execution times. On the other hand, hybrid alternatives excel in the execution time dimension, offering a drastic reduction. However, they are outperformed in terms of economic cost and result quality.

For a more detailed quantitative analysis, in \textit{Figures} from~\ref{fig:wfhs-cost-comparison} to ~\ref{fig:wfhs-quantumerror-comparison} a comparative box-plot visualization of the four studied dimensions is provided. In terms of cost, classical configurations shows a minimal cost profile, represented by a nearly flat boxplot at the lower bound of scale. On the other hand, hybrid configurations shows a significant value of cost higher thant the classical ones. Additionally, the hybrid box suggests that the pricing is sensitive to the quantum machine selected because of the large interquartile range. Comparing the execution time, hybrid configurations exhibit a performance advantage, reducing the median execution time to approximately 1500 ns. Conversely, classical solutions operate at a significantly slower baseline, clustering tightly just above the 5000 ns mark. Notably, unlike the other metrics, reliability does not serve as a clear discriminating factor between clusters; rather, it appears homogeneously distributed throughout the entire design space. This statement is illustrated in \textit{Figure}~\ref{fig:wfhs-reliability-comparison}.

This global characterization of the design space does not, in itself, resolve the architectural choice; rather, the suitability of a solution fluctuates according to the temporal constraints imposed by the context. Subsequently, the global utility of the generated configurations is evaluated under two opposing weather scenarios, modeling how the urgency of obtaining a weather forecast serves as a determining factor.

\paragraph{\textbf{5.3.2.2 Context-Aware Decision Making}}
\paragraph{\textbf{Normal weather forecast}} The first established context models the generation of conventional weather forecasts. In this scenario, execution time is not a priority, therefore, solutions with high quality and low monetary cost are sought. To achieve this, weights are assigned within the utility function to increase high solution quality and low monetary cost ($w_{cost} = 0.5; w_{executiontime}= 0;$$w_{reliability}= 0; w_{quantumerror}= 0.5$).

\begin{figure}[ht]
    \centering
    \includegraphics[width=0.75\linewidth]{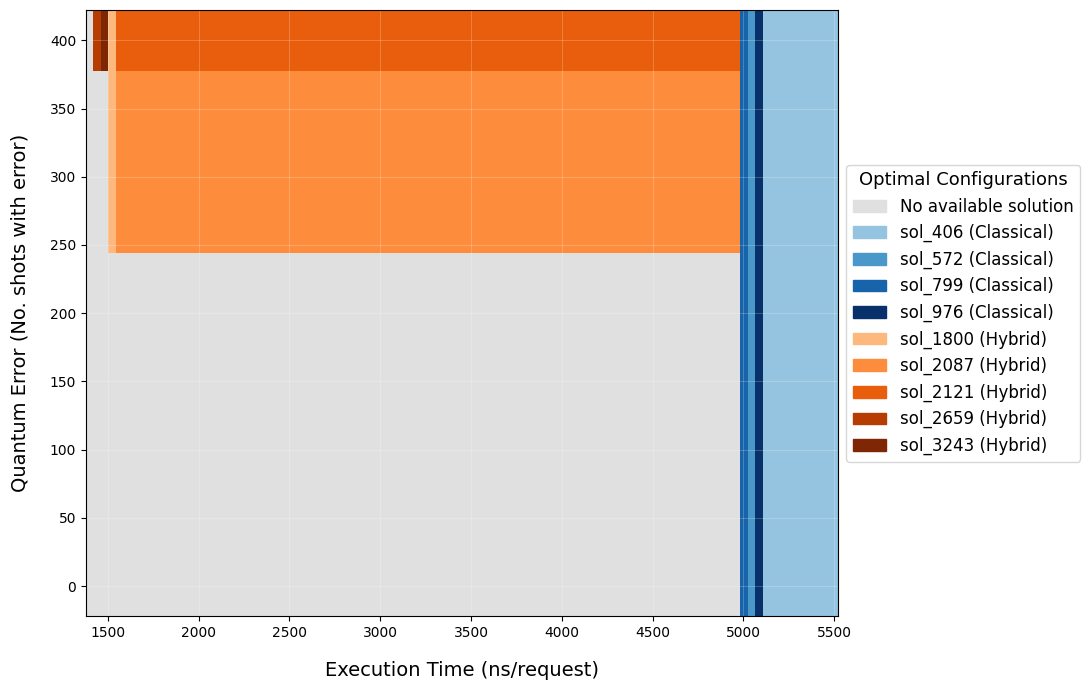}
    \Description{Distribution of optimal solutions among quantum error and execution time dimension of the Normal weather forecast scenario.}
    \caption{Normal Forecast Scenario - Optimal Solutions}
    \label{fig:wfhs_normal_optimal}
\end{figure}

Under these conditions, the design space of optimal solutions points out classical configurations. Consequently, their ability to deliver high-quality predictions at a low cost maximizes global utility. Conversely, hybrid solutions, despite their low execution times, are relegated due to penalties incurred in the quality and cost dimensions, yielding marginal utility values in this context. Table~\ref{tab:wfhs_scenario_1} lists the top 5 solutions, ranked from highest to lowest utility.
However, \textit{Figure~\ref{fig:wfhs_normal_optimal}} reveals the comprehensive scope of the design space for the weather forecast scenario. In this regard, the heatmap demonstrates that optimality is not absolute but contingent upon the constraints imposed on the system. The plot clearly delineates the decision boundaries where the system transitions toward hybrid configurations (in orange) or alternates between distinct classical architectures to maximize the utility function.

\begin{table}[H]
\caption{Normal weather forecast scenario - Top 5 solutions}\label{tab:wfhs_scenario_1}
\begin{center}
\begin{tabular}{|c|c|c|c|c|}
\hline
\textbf{Solution} & 
\textbf{Cost ($\$/req$)} & 
\textbf{\makecell{Execution Time\\($\mathrm{ns}/req$)}} &  
\textbf{Reliability ($\%$)} &  \textbf{\makecell{Quantum Error\\$(No.~shots~with ~error/req)$}}
\\
\hline
sol-406 & 0.435501 & 5120.023297 & 86.837789 & 0.0
\\
\hline
sol-976 & 0.435501 & 5084.576769 & 83.122861 & 0.0
\\
\hline
sol-1117 & 0.435806 & 5045.983657 & 79.061943 & 0.0
\\
\hline
sol-548 & 0.435806 & 5084.023228 & 82.951582 & 0.0
\\
\hline
sol-370 & 0.437051 & 5115.606844 & 86.629120 & 0.0
\\
\hline
\end{tabular}
\end{center}
\end{table}

\paragraph{\textbf{Emergency Weather Forecast}} In this second context, an alert situation is modeled. Given this scenario, priorities are drastically inverted, assigning greater weight to execution time and imposing a stricter temporal constraint (between 1000 and 2000 ns/request). Consequently, classical weather forecast configurations fail to meet the established restriction, rendering their utility zero in this scenario.

\begin{figure}[ht]
    \centering
    \includegraphics[width=0.75\linewidth]{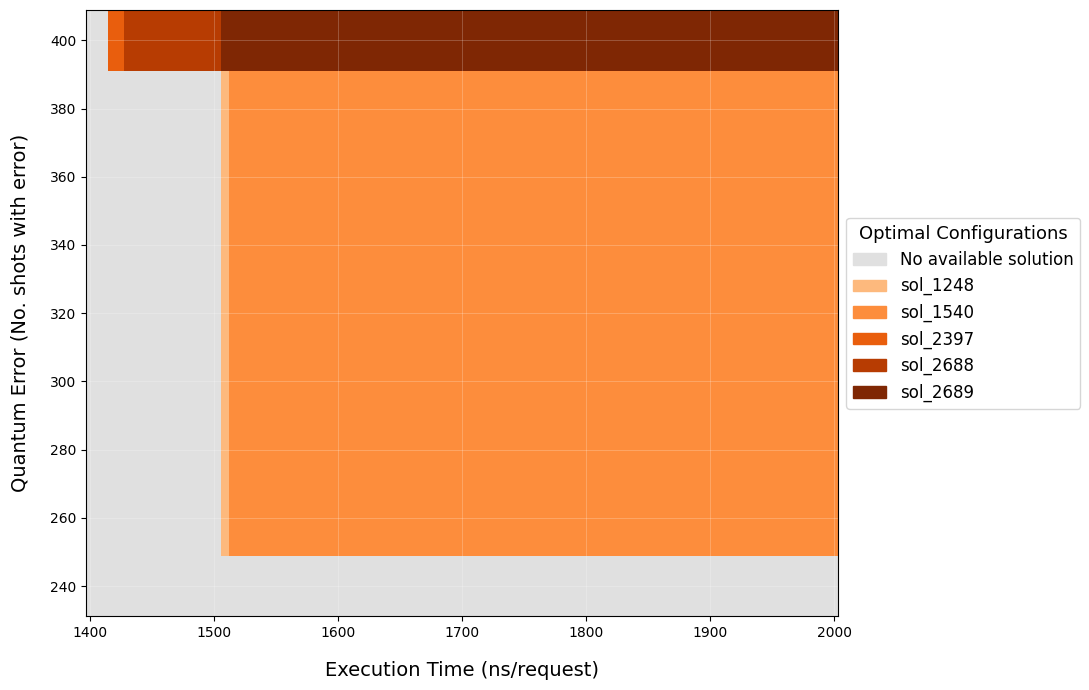}
    \Description{Distribution of optimal solutions among quantum error and execution time dimension of the emergency weather forecast scenario.}
    \caption{Emergency Forecast Scenario - Optimal Solutions}
    \label{fig:wfhs_emergency_optimal}
\end{figure}

In this context, hybrid configurations emerge as the only viable alternative satisfying the defined bounds. Furthermore, within the set of hybrid configurations, it is necessary to determine the most suitable architecture. To this end, \textit{Table~\ref{tab:wfhs_scenario_2}} lists the top 5 solutions for this scenario, seeking to balance cost, reliability, and precision against execution time. Additionally, \textit{Figure~\ref{fig:wfhs_emergency_optimal}} illustrates the evolution of the optimal solution within the design space constrained by execution time and precision. Notably, the tonal variations denote transitions between distinct hybrid architecture configurations, demonstrating the system's capability to identify the best possible configuration in response to varying Quality of Service (QoS) requirements.

\begin{table}[H]
\caption{Emergency weather forecast scenario - Top 5 solutions}\label{tab:wfhs_scenario_2}
\begin{center}
\begin{tabular}{|c|c|c|c|c|}
\hline
\textbf{Solution} & 
\textbf{Cost ($\$/req$)} & 
\textbf{\makecell{Execution Time\\($\mathrm{ns}/req$)}} &  
\textbf{Reliability ($\%$)} &  \textbf{\makecell{Quantum Error\\$(No.~shots~with ~error/req)$}}
\\
\hline
sol-2689 & 14.072453 & 1416.442747 & 86.690606 & 385.249940
\\
\hline
sol-2682 & 14.031743 & 1417.176797 & 86.765443 & 385.249943
\\
\hline
sol-2729 & 14.077542 & 1416.432423 & 86.615533 & 385.249936
\\
\hline
sol-2666 & 14.036831 & 1417.166631 & 86.690606 & 385.249940
\\
\hline
sol-2688 & 14.071235 & 1416.547517 & 86.569894 & 385.249934
\\
\hline
\end{tabular}
\end{center}
\end{table}

\subsection{Discussion}
\label{subsec:discussion}
Following an exhaustive analysis of the results obtained for both case studies, the subsequent section details how the proposed framework addresses the formulated research questions.

\subsubsection{RQ1. How can we automatically identify the set of configurations that satisfy the architectural constraints
characteristic of quantum-classical service-based
systems?}
The proposed system enables the identification of the set of configurations that satisfy the architectural constraints characteristic of quantum-classical service-based systems. To achieve this, the system provides an analytical method that delimits the decision boundaries between different deployment types, depending on the established set of preferences. This is accomplished through the generation of a heatmap, which visualizes the set of constraints for which a valid solution (either classical or quantum) exists, as well as identifying the design space where the generated configurations fail to meet the requirements. Furthermore, the analysis method ensures compliance with SLA criteria even before considering the utility of the candidate solutions.

\subsubsection{RQ2. How can we analyze the trade-offs among relevant
non-functional system properties (e.g., cost, performance, reliability) in quantum-
classical service-based systems across the architectural design space
and under prescribed levels of uncertainty?}
Through the established analysis process, it is possible to examine the design trade-off space for a given hybrid application. In this analysis, a utility calculation is performed for each generated solution across multiple defined scenarios. Consequently, it is possible to identify a configuration that maximizes utility by balancing conflicting non-functional properties, including critical factors such as uncertainty and economic cost.

Furthermore, it is noteworthy that this method integrates a specific characteristic of hybrid systems, such as the number of erroneous shots (quantumerror). This establishes the groundwork for integrating additional dimensions specific to quantum systems, such as queueing delays or circuit decoherence levels, among others.

In summary, this procedure enables the exploration of the architectural design space by revealing the decision boundaries for the selection of the optimal configuration, depending on the software architect's Quality of Service requirements.

%% file: 6_ThreatsToValidity/threatstovalidity.tex
In this section, the main threats to validity are discussed in order to ensure that the previously presented results are as reliable as possible through their identification and analysis. The structure follows the Internal, External, Construct and Conclusion subsections.

\subsection{Internal Validity}
One of the internal threats faced has been obtaining information about the reliability of digital computing machines from the selected provider, Amazon Web Services (AWS)~\footnote{\url{https://aws.amazon.com/}}. This provider promises 99.5\% availability for its machines in its Service Level Agreement (SLA)~\footnote{\url{https://aws.amazon.com/compute/sla/}}. However, some studies, such as the work by Zhenyu Wen et al.~\cite{7435330} and Peter Garraghan et al.~\cite{6754595}, have shown that this availability percentage is not exactly accurate, as availability can decrease by \textit{2.46\%}, dropping to 97.54\%. Therefore, based on the information provided in these works and the absence of this type of information from service providers, it was decided to use a range of availability values from 97.54\% (the minimum availability percentage) to 99.5\% (the value promised in the SLA Contract). The assignment of availability values was done using a random distribution to assign a different probability to each of the digital computing machines used in this study. It is acknowledged that these values are arbitrary. However, the main objective of this article is to provide validity for the use of these techniques in generating deployment configurations for hybrid applications.

On the other hand, the need to manage computational complexity arises. Given the variability of computing machines and the number of services, the number of configurations to be generated has been intentionally limited to reduce the combinatorial explosion. This decision, although it reduces the exhaustiveness of the work, was necessary to address the study. This is evidenced when observing the times and number of configurations generated in both use cases (\textit{Table~\ref{tab:haiq_results}}), even though the classical and quantum machines were deliberately limited.

Additionally, a notable limitation in the simulation of the quantum environment is presented. In this study, the real operational conditions of QPUs have not been implemented. Specifically, this paper does not consider queue times, which are, in some cases, too long due to high demand and limited availability of quantum resources. In the same way, the restricted availability windows of the quantum machines have not been considered yet. These factors significantly influence the deployment performance and will be integrated into the future evolution of the proposed system.

\subsection{External Validity}
Regarding external validity, several threats arise from the current state of quantum hardware and its providers. Although AWS offers access to quantum computing machines, the metrics provided are not suitable for studying the performance of hybrid configurations. A significant metric for this is the CLOPS attribute, explained in \textit{Section\ref{sec:background}}. For this reason, International Business Machines (IBM)~\footnote{\url{https://www.ibm.com/}} was used as the provider for the QPUs. However, since the cost of IBM machines is the same for each computer, in order to achieve greater richness in the trade-offs spaces, the quantum computer provider IQM~\footnote{\url{https://meetiqm.com/}} was also considered. IQM machines include the CLOPS metric, as well as a payment quota dimensionality similar (0.30\$/s~\footnote{\url{https://meetiqm.com/products/iqm-resonance/}}) to that of IBM (96\$/minute~\footnote{\url{https://www.ibm.com/quantum/products}}).
This lack of standardisation in hardware metrics between different QPU providers poses a challenge, as quantum software development tools must be adapted to each provider’s specific metrics rather than relying on a universal abstraction layer. 
Furthermore, in this work, only gate-based quantum computers have been considered. There are several reasons for this decision. At first, they are the most widely used quantum computers, as well as universal computers~\cite{DBLP:books/sp/Hidary21}, meaning they can execute any algorithm. Secondly, this is an incipient research study, so it was decided to limit the scope to gate-based quantum computers. Including all types of architectures would complicate the problem's definition and proper delimitation in this initial phase.

\subsection{Construct Validity}
In this section, a threat to the study's validity linked to the proposed architectural style is identified. Given the incipient nature of the hybrid (classical-quantum) application architecture concept, there is a risk that the constraints defined for the service or deployment do not fully reflect optimal operational conditions.
This design limitation is also present in the execution time assigned to tasks. As this is an initial study, the quantum algorithm being simulated has not been addressed in detail. Consequently, the execution time defined in the specification is not based on any specific algorithm or even on the execution times of similar algorithms. This simplification was necessary to scope and address the problem in this initial phase, but its modification will be required for validation with real use cases.

\subsection{Conclusion Validity}
Potential threats to conclusion validity arise from the fact that quantum computing is an incipient field. The observed trade-offs, while sufficient for validating the technique, may change as quantum technologies reach a higher degree of maturity. Likewise, these initial findings could be altered when real-world hybrid (quantum-classical) use cases are introduced.

%% file: 7_RelatedWork/relatedwork.tex
Related approaches can be categorized into: (i)~Formalization and quantitative analysis/optimization of architectures, (ii)~architecture of quantum computing systems, and (iii)~placement/deployment of complex software systems.

\paragraph{\textbf{Formalization and quantitative analysis/optimization of architectures}}
There is extensive related work on model-based performance prediction~\cite{DBLP:journals/tse/BalsamoMIS04} and optimization of quantitative aspects of architectures~\cite{DBLP:journals/jss/GrunskeA13} that typically use mechanisms like stochastic search and/or Pareto analysis~\cite{DBLP:conf/icse/EsfahaniMR13,5069138,DBLP:conf/qosa/MeedeniyaMAG11,Martens:2010:AIS:1712605.1712624,Bondarev:2007:EPT:1216993.1217020,DBLP:journals/jss/BeckerKR09,DBLP:journals/tse/BroschKBR12}.
{\em PerOpteryx}~\cite{Martens:2010:AIS:1712605.1712624} takes as input an architectural model described using the Palladio component model and tries to automatically improve it by searching for pareto-optimal solutions employing a genetic algorithm. 
{\em ArcheOpterix}~\cite{5069138} uses an evolutionary algorithm for optimizing the architecture of embedded systems specified in AADL~\cite{feiler2004overview}.
{\em DeepCompass}~\cite{Bondarev:2007:EPT:1216993.1217020} is a framework that analyzes different architectural alternatives along the dimensions of performance and cost to find pareto-optimal solutions. 
While these approaches can optimize quantitative aspects of designs they do not support synthesis of configurations. 
Other approaches~\cite{DBLP:conf/itng/DwivediGPS14,DBLP:conf/icse/BagheriTS14} combine structural synthesis with simulation and dynamic analysis to provide estimates of quantitative properties of design variants.
{\em TradeMaker}~\cite{DBLP:conf/icse/BagheriTS14} synthesizes design spaces for relational database mappings, in which individual designs are subject to static and dynamic analysis to extract performance metrics. Dwivedi et al.~\cite{DBLP:conf/itng/DwivediGPS14} propose using architectural models coupled with automated design space generation for making fidelity and timeliness tradeoffs.
These approaches share with ours the idea of synthesizing a solution space from a set of constraints and analyzing individual solutions independently. However, they are not equipped to provide quantitative guarantees under uncertainty, which rely on checking sophisticated properties (typically encoded as temporal logic formulas) via numerical methods and exhaustive state space exploration techniques.

The class of technique we employ to support our approach~\cite{DBLP:conf/icse/Camara20}, which combines the capabilities of quantitative verification with configuration synthesis, has been applied to the analysis of architectural design trade-off spaces in contexts such as security countermeasure selection~\cite{skandylas2022security}, as well as pub-sub and service-based systems~\cite{DBLP:journals/jss/CamaraGS19}. 
However, such proposals do not consider the idiosyncrasy of hybrid quantum-classical systems, including aspects of the problem related to machine-algorithm placement, which are explicitly addressed in our proposed solution and captured in our formalization of hybrid quantum-classical architectural style.

\paragraph{\textbf{Quantum Service-Oriented Architecture}} In terms of the Quantum Service-Oriented Architecture (QSOA), research papers are beginning to emerge that propose Service-Oriented Computing as a way of integrating quantum computing into current software solutions.

On the one hand, we highlight notable works that address issues of adapting tools and processes for the development of hybrid services. Among these, notable works include Alvarado et al.\cite{alvarado2022guide}, which presents a guide for converting quantum circuits into web services, and Nguyen et al.\cite{nguyen2024qfaas}, which introduces the concept of \emph{quantum-function-as-a-service}, where a serverless model is used to execute quantum circuits.

Another line of research in quantum services focuses on their deployment. Noteworthy here is the work of Karoline Wild et al.\cite{9233151}, which introduces TOSCAQ a model featuring two deployment styles based on the Topology and Orchestration Specification for Cloud Applications (TOSCA) standard to automate the deployment and orchestration of quantum applications. Additionally, Alvarez et al.\cite{romero2024enabling} propose a strategy for the creation and continuous deployment of quantum services using an OpenAPI extension. 

On the other hand, there are also works that address the modelling challenges of quantum or hybrid software. In this context, the work of Carlos et al.\cite{DBLP:books/sp/22/Perez-Delgado22} introduces Q-UML, an extension to classical UML, designed for the structural and behavioural representation of quantum search algorithms. Similarly, Pérez-Castillo and Piattini\cite{DBLP:journals/computing/Perez-CastilloP22} explore how UML can be adapted to facilitate the co-design of classical-quantum systems, enabling the seamless integration of quantum functionalities within traditional software architectures.

Beyond UML-based approaches, constraints and architectural modelling languages play a crucial role in defining quantum software architectures. Medvidovic et al.\cite{DBLP:journals/tosem/MedvidovicRRR02} investigate how software architectures can be represented within UML, providing insights into the application of architectural constraints in quantum systems.

\paragraph{\textbf{Placement/Deployment of complex software systems}}
The deployment of complex software systems is a type of problem that has already been addressed in classical computing, with numerous studies examining it from various perspectives. Among these studies, we can highlight the work of Herrera et al.~\cite{Herrera2023}, which provides recommendations based on the composition of service applications and user requirements; the study by Yang Hu et al.~\cite{Hu2019}, which seeks to optimize the deployment of microservice-based applications in cloud environments by minimizing inter-component traffic; and the work of Selimi et al.~\cite{Selimi2017}, which aims to leverage network state information to support service placement decisions. Other studies focus on the economic cost aspect of this deployment. This is the case in Olivier Belli’s work~\cite{7830663}, where they implement a cost-optimization tool through a resource consumption model (RCM), enabling users to automatically compare the prices of different cloud infrastructure providers for their applications. Another example is the study by Ang Li et al.~\cite{10.1145/1879141.1879143}, which proposes an evaluation tool that, through controlled testing across different providers, estimates and compares the performance and costs of each, helping users select the most suitable provider.

Furthermore, some studies adopt a metaheuristic approach. In this vein, the work by Hemant Kumar Apat et al.~\cite{APAT2024100379} is noteworthy, as it proposes a hybrid metaheuristic algorithm for IoT service placement based on multiple objectives, optimizing parameters such as makespan, cost, and energy consumption. There is also the study by Poria Pirozman et al.~\cite{Pirozmand}, which proposes using a genetic algorithm and energy-aware scheduling heuristic (GAECS) to address the multi-objective task scheduling problem in cloud computing, aiming to reduce both makespan and energy consumption.

In the Quantum Computing study area, works addressing the challenge of selecting the most suitable quantum machine for a specific quantum circuit are starting to appear. It is the case of NISQ Analizer~\cite{DBLP:conf/summersoc/SalmBBLWW20} where a suggestion of the optimal implementation of an algorithm and the appropriate quantum computer for certain input data is made or the work of Nils Quetschlich et al.~\cite{10.1145/3673241} that propose MQT Predictor, a methodology for the automatic choice of a quantum computer for an application as well as the best compiler for the selected machine.

As can be observed, numerous studies, like this one, seek to address the \emph{placement problem} by applying different techniques with diverse final objectives. However, these approaches do not consider the application of quantum computing or its characteristics to achieve an optimal deployment of hybrid (quantum-classical) applications.

%% file: 8_ConclusionAndFutureWork/conclusionandfuturework.tex
As NISQ computer companies rapidly advance in developing fault-tolerant quantum computers, efforts must begin to emerge in the field of quantum software engineering in order to bridge the gap between experimental devices and practical applications. Addressing this challenge, this work aims to pave the way for integrating quantum components into Service-Oriented Architecture. For this purpose, a trade-off decision tool for hybrid (quantum-classical) applications is presented.
The evaluation of the proposed framework has been done by analysing the Hybrid Search Application and the Weather Hybrid Forecast System examples. Within this evaluation, clear answers to the formulated research questions have been provided.
Regarding the automatic identification of valid configurations (\textbf{\textit{RQ1}}), results demonstrate that the proposed procedures enable the identification of diverse regions within the design trade-off space. By utilizing the decision region map, it is possible to provide software architects with a clear visualization of the best configuration, depending on the established constraints for execution time and quantum error. On the other hand, concerning the trade-off analysis under uncertainty (\textbf{\textit{RQ2}}), results reveal that the optimal configuration is inherently context-dependent. This characteristic renders the proposed framework a dynamic decision-making tool that prioritizes classical solutions when economic constraints are dominant, while favoring quantum configurations in high-performance scenarios where execution time is critical.

Consequently, this work presents a context-aware framework that enables the identification of optimal deployments for a given hybrid application through the use of a utility function. The novelty of this approach lies in the integration of quantum components as a valid deployment alternative to existing classical services, the explicit consideration of hardware specifications for both quantum and classical machines, the generation of structurally correct deployment configurations, and the inherent dependency of service quality attributes on the specific hardware-service mapping thereby providing a mechanism to manage this complexity of future hybrid systems.

In future work, several validity threats will be addressed to develop a more complete and sophisticated prototype. For instance, it is planning to simulate the queuing behavior of quantum computers and incorporate software quality metrics to account for the stochastic behavior of the service applications. In the long term, it is necessary to develop this methodology to support the deployment of hybrid applications across the entire spectrum of computing (Cloud-Fog-Edge). Additionally, the creation of a fully functional proof of concept for integration into a practical deployment tool that will enable optimal deployment of a hybrid (quantum-classical) application has started to be developed. Currently, only cloud machine providers are considered, but the plan is to extend this to analyze applications deployed across the computing continuum, which will introduce new challenges and further quality of service metrics to evaluate.